\documentclass[11pt]{article}       
\usepackage{geometry}               
\usepackage{amsmath,amsfonts,mathrsfs,amssymb,amscd,amsthm}
\usepackage{graphicx}
\usepackage{hyperref}
\usepackage{textcomp}
\usepackage[utf8]{inputenc} 
\usepackage[T1]{fontenc}    
\usepackage{url}            
\usepackage{booktabs}       
\usepackage{xfrac}       
\usepackage{microtype}      
\usepackage{mathtools}
\usepackage{mathrsfs}
\usepackage{bm}
\usepackage{bbm}
\usepackage{cleveref}
\usepackage{placeins}
\usepackage{multirow}
\usepackage{multicol}
\usepackage{array}
\usepackage{caption}
\usepackage{subcaption}
\usepackage{rotating}
\usepackage{makecell} 
\usepackage{marginnote}
\usepackage{svg}
\usepackage{cite}
\usepackage{algorithm}
\usepackage{algpseudocode}
\usepackage{accents}

\newtheorem{proposition}{Proposition}

\newtheorem{assumption}{Assumption}
\newtheorem{corollary}{Corollary}
\newtheorem{lemma}{Lemma}
\newtheorem{theorem}{Theorem}
\newtheorem{remark}{Remark}

\newcommand{\xpost}{\bar{x}}
\newcommand{\xprior}{\bar{x}^-}

\newcommand{\Xpost}{\Sigma_{x,t}}

\newcommand{\Xprior}{\Sigma_{x,t}^-}
\newcommand{\Xpriorlowinit}{\underline{\Sigma}_{x,0}^{-}}
\newcommand{\Xpriorhighinit}{\overline{\Sigma}_{x,0}^{-}}

\newcommand{\Xpriorlow}{\underline{\Sigma}_{x,t}^{-}}
\newcommand{\Xpriorhigh}{\overline{\Sigma}_{x,t}^{-}}
\newcommand{\Xpriorlowss}{\underline{\Sigma}_{x,\infty}^{-}}
\newcommand{\Xpriorhighss}{\overline{\Sigma}_{x,\infty}^{-}}
\newcommand{\Xpostlow}{\underline{\Sigma}_{x,t}}
\newcommand{\Xposthigh}{\overline{\Sigma}_{x,t}}
\newcommand{\Xpostlowss}{\underline{\Sigma}_{x,\infty}}
\newcommand{\Xposthighss}{\overline{\Sigma}_{x,\infty}}

\newcommand{\Xpostss}{\Sigma_{x,\infty}}
\newcommand{\Xpriorss}{\Sigma_{x,\infty}^-}

\newcommand{\Xpostssopt}{\Sigma_{x,\infty}^*}
\newcommand{\Xpriorssopt}{\Sigma_{x,\infty}^{-,*}}

\newcommand{\Xpostinit}{\Sigma_{x,0}}
\newcommand{\Xpriorinit}{\Sigma_{x,0}^-}

\newcommand{\Xpriord}{\Sigma_{x,t}^{-,d}}
\newcommand{\Xpriornextd}{\Sigma_{x,t+1}^{-,d}}

\newcommand{\Xpriornext}{\Sigma_{x,t+1}^-}
\newcommand{\Xpriorinitopt}{\Sigma_{x,0}^{-,*}}

\newcommand{\Xpostprev}{\Sigma_{x,t-1}}
\newcommand{\xnom}{\hat{x}_0^-}
\newcommand{\Xnom}{\hat{\Sigma}_{x,0}^-}
\newcommand{\Pdist}{\mathbb{P}}
\newcommand{\Qdist}{\mathbb{Q}}
\newcommand{\Qhat}{\hat{\mathbb{Q}}}
\newcommand{\ambset}{\mathbb{D}}
\newcommand{\Gauss}{\mathcal{N}}

\newcommand{\Priordistinit}{\Pdist_{x,0}^-}
\newcommand{\Postdistprev}{\Pdist_{x,t-1}}

\newcommand{\Postdistopt}{\Pdist_{x,t}^*}
\newcommand{\Priordistopt}{\Pdist_{x,t}^{-,*}}
\newcommand{\Priordistinitopt}{\Pdist_{x,0}^{-,*}}

\newcommand{\lammin}[1]{\lambda_{\min}(#1)}
\newcommand{\lammax}[1]{\lambda_{\max}(#1)}

\newcommand{\Bures}{\mathcal{B}}
\newcommand{\Wass}[2]{W_2(#1, #2)}
\newcommand{\Gelb}[2]{G(#1, #2)}
\newcommand{\real}[1]{\mathbb{R}^{#1}}
\newcommand{\symm}[1]{\mathbb{S}^{#1}}
\newcommand{\psd}[1]{\symm{#1}_{+}}
\newcommand{\pd}[1]{\symm{#1}_{++}}

\newcommand{\mref}[1]{\textbf{(\ref{#1})}} 
\DeclareMathOperator{\blkdiag}{blkdiag}

\newenvironment{methodlist}{%
  \begingroup
  \setlength\itemsep{2pt}%
  \begin{enumerate}%
}{%
  \end{enumerate}%
  \endgroup
}

\DeclareMathOperator{\Tr}{Tr}
\Crefname{figure}{Fig.}{Figs.}
\Crefname{remark}{Remark}{Remarks}
\crefname{assumption}{Assumption}{Assumptions}

\title{Distributionally Robust Kalman Filter\thanks{This work was supported in part by the Information and Communications Technology Planning and Evaluation
(IITP) grant funded by MSIT(2022-0-00124, 2022-0-00480). The work of A. Hakobyan was supported in part by the Higher Education and Science Committee of RA (Research project 24IRF-2B002).} }%
\author{Minhyuk Jang,  Astghik Hakobyan, and Insoon Yang\thanks{The first two authors contributed equally. A preliminary version of this paper was presented at the 2025 IEEE Conference on Decision and Control~\cite{jang2025steady}.}\thanks{M. Jang is with the Department of Mechanical Science and Engineering, Grainger College of Engineering, University of Illinois Urbana-Champaign, Urbana, IL, United States {\tt\small jang64@illinois.edu}. A. Hakobyan is with the Center for Scientific Innovation and Education and National Polytechnic University of Armenia, Yerevan, Armenia {\tt\small astghik.hakobyan@csie.am}. I. Yang is with 
the Department of Electrical and Computer Engineering and ASRI, Seoul National University, Seoul, South Korea  {\tt\small insoonyang@snu.ac.kr.}}}

\date{}

\begin{document}

\maketitle

\begin{abstract}                          
We study state estimation for discrete-time linear stochastic systems under distributional ambiguity in the initial state, process noise, and measurement noise. We propose a noise-centric distributionally robust Kalman filter (DRKF) based on Wasserstein ambiguity sets imposed directly on these distributions. 
This formulation excludes dynamically unreachable priors and yields a Kalman-type recursion driven by least-favorable covariances computed via semidefinite programs (SDP).
In the time-invariant case, the steady-state DRKF is obtained from a \emph{single} stationary SDP, producing a constant gain with Kalman-level online complexity.
We establish the convergence of the DR Riccati covariance iteration to the stationary SDP solution, together with an explicit sufficient condition for a prescribed convergence rate.
We further show that the proposed noise-centric model induces \emph{a~priori} spectral bounds on all feasible covariances and a Kalman filter sandwiching property for the DRKF covariances.
Finally, we prove that the steady-state error dynamics are Schur stable, and the steady-state DRKF is asymptotically minimax optimal with respect to worst-case mean-square error.
\end{abstract}

\section{Introduction}

State estimation is a fundamental problem in systems and control, where the objective is to infer the system state from noisy and partial measurements. 
For linear systems with Gaussian noise and known statistics, the Kalman filter (KF)~\cite{Kalman1960} provides the optimal minimum mean-square error (MMSE) estimator.
In practice, however, noise statistics are rarely known \emph{a~priori}: they are typically learned from limited data, subject to modeling error, and may vary over time.
Even moderate misspecification can significantly degrade estimation performance, while severe mismatch may lead to divergence.

 To mitigate this sensitivity, several robust estimation frameworks have been developed.
The $H_{\infty}$ filter~\cite{simon2006optimal} guarantees bounded estimation error under worst-case disturbances, but it is often conservative in stochastic settings.
Risk-sensitive filters (e.g.,~\cite{speyer1992optimal} and~\cite{whittle1981risk}) penalize large errors exponentially via an entropic risk measure, but their performance relies on an accurate specification of the underlying noise distributions, which may be difficult to justify in practice.

In contrast, recent advances in distributionally robust optimization (DRO) have motivated growing interest in \emph{distributionally robust state estimation} (DRSE).
In DRSE, the true noise distributions are assumed to lie in an ambiguity set centered at a nominal model, and the estimator minimizes the worst-case expected error over this set.
While distributional robustness has been extensively studied in control design
(e.g.,~\cite{van2015distributionally,yang2020wasserstein,schuurmans2023general,li2024tac,hakobyan2024wasserstein,mcallister2024distributionally,brouillon2025distributionally}),
its application to state estimation remains comparatively limited.

Existing DRSE formulations vary along two key dimensions: 
(i) the \emph{geometry} of the ambiguity set, such as $\phi$-divergence balls, moment-based sets~\cite{wang2021robust,wang2021distributionally,zorzi2016robust,levy2012robust}, or Wasserstein balls~\cite{NEURIPS_DRKF,nguyen2023bridging,lotidis2023wasserstein,han2024distributionally,kargin2024distributionally};
and (ii) the \emph{location} where ambiguity is imposed, e.g., on the joint state--measurement distribution~\cite{NEURIPS_DRKF}, on the prior state and measurement noise~\cite{nguyen2023bridging}, or via bicausal optimal transport models that encode temporal dependence~\cite{han2024distributionally}.
These choices have significant implications for tractability, interpretability, and long-term behavior.

Despite this progress, \emph{steady-state} DRSE remains underdeveloped.
For long-horizon operation, constant-gain estimators with fixed offline complexity are essential.
However, many existing time-varying DRKFs require solving an optimization problem at each time step, or lead to horizon-dependent formulations whose size grows with time.
While the frequency-domain minimax approach of~\cite{kargin2024distributionally} characterizes an optimal linear time-invariant estimator, the resulting filter is generally non-rational and does not admit a finite-dimensional state-space realization.

The preliminary version~\cite{jang2025steady} of this work derived a steady-state DRKF by placing ambiguity sets on the prior state and measurement noise distributions.
In this paper, we substantially extend this line of work by adopting a \emph{noise-centric} formulation, in which Wasserstein ambiguity sets are imposed directly on the initial state, process noise, and measurement noise distributions.
This modeling choice enforces a form of \emph{dynamic consistency}: all admissible prior distributions arise through propagation of the system dynamics under admissible noise realizations.
Moreover, it admits a finite-dimensional steady-state characterization and yields a constant-gain DRKF from a \emph{single} stationary semidefinite program (SDP), while providing several theoretical and computational advantages.

Our main contributions are summarized as follows.

\begin{itemize}

\item \textbf{Noise-centric ambiguity modeling.}
We define Wasserstein ambiguity sets directly on the initial state, process noise, and measurement noise distributions.
This modeling choice enforces \emph{dynamic consistency}, ensuring that all admissible prior distributions arise through propagation of the system dynamics under admissible noise realizations, and forms the structural basis for the subsequent theoretical and algorithmic developments.

\item \textbf{Steady-state DRKF via a single stationary SDP.}
In the time-invariant setting, we obtain a constant-gain steady-state DRKF by solving a \emph{single} stationary SDP offline, yielding Kalman-level per-step online complexity.

 \item \textbf{Convergence with explicit rate conditions.}
We establish the convergence of the DR Riccati iteration governing the state-error covariance to a unique steady-state solution characterized by the stationary SDP.
Moreover, we derive an explicit sufficient condition on the ambiguity radii that guarantees a \emph{prescribed} contraction rate.

 \item \textbf{Spectral structure and steady-state guarantees.}
We show that Wasserstein ambiguity induces explicit \emph{a~priori} eigenvalue bounds (an \emph{eigenvalue tube}) and a KF-sandwich property for the DRKF covariances.
We further establish Schur stability of the steady-state estimation error dynamics and asymptotic minimax optimality with respect to the worst-case mean-square error (MSE).

\end{itemize}

The remainder of the paper is organized as follows.
\Cref{sec:prelim} introduces the problem formulation and Wasserstein ambiguity sets.
\Cref{sec:drkf} presents the finite-horizon and steady-state DRKF constructions and their SDP reformulations.
\Cref{sec:theoretical} develops spectral boundedness, convergence, stability, and optimality results.
\Cref{sec:simulations} illustrates the performance of the proposed methods through numerical experiments.

\section{Problem Setup}\label{sec:prelim}

\noindent{\bf Notation.}
Let $\symm{n}$ denote the set of symmetric matrices in $\real{n\times n}$, and let
$\psd{n}$ and $\pd{n}$ be the positive semidefinite and positive definite cones.
For $A,B\in\symm{n}$, write $A\succeq B$ (resp.\ $A\succ B$) if $A-B\in\psd{n}$ (resp.\ $\pd{n}$).
For a matrix $A$, $\|A\|_2$ and $\|A\|_F$ denote the spectral and Frobenius norms. 
For $A\in\symm{n}$, $\lammin{A}$ and $\lammax{A}$ denote its smallest and largest eigenvalues, while $\sigma_{\min}(A)$ and $\sigma_{\max}(A)$ denote its smallest and largest singular values. 
Let $\mathcal{P}(\mathcal{W})$ be the set of Borel probability measures supported on $\mathcal{W}\subseteq\real{n}$,
and let $\mathcal{P}_2(\mathcal{W})$ be those with finite second moments.
For  probability measures $\Pdist,\Qdist$, their product measure is denoted by $\Pdist\otimes\Qdist$.  Let $\mathbf{1}_N\in\mathbb{R}^N$ denote the vector of all ones, and let $\blkdiag(A_1,\dots,A_N)$ denote the block-diagonal matrix with diagonal blocks $A_1,\dots,A_N$.

\subsection{Distributionally Robust State Estimation Problem}

Consider the discrete-time linear stochastic system\footnote{Throughout, we assume that the system matrices $A_t$ and $C_t$ are known, and we focus
exclusively on uncertainty in the noise statistics.}
\begin{equation}\label{eqn:system_eq}
\begin{split}
    x_{t+1} &= A_t x_t + w_t,\\
    y_t &= C_t x_t + v_t,
\end{split}
\end{equation}
where $x_t \in \real{n_x}$ and $y_t \in \real{n_y}$ denote the system state and output. 
The process noise $w_t \in \real{n_x}$ and  measurement noise $v_t \in \real{n_y}$ follow distributions $\Qdist_{w,t} \in \mathcal{P}(\real{n_x})$ and $\Qdist_{v,t} \in \mathcal{P}(\real{n_y})$, while the initial state $x_0$ follows $\Qdist_{x,0} \in \mathcal{P}(\real{n_x})$. 
We assume that $x_0$, $\{w_t\}$, and $\{v_t\}$ are mutually independent, and that each of $\{w_t\}$ and $\{v_t\}$ is temporally independent.\footnote{Such independence assumptions are standard in the control and estimation literature; see, e.g., \cite[Sect.~2.2]{anderson2005optimal} and \cite[Sect.~5.1]{simon2006optimal}.}

At each time $t \ge 0$, the estimator has access to the measurement history $\mathcal{Y}_t \coloneqq \{y_0, \dots, y_t\}$. 
The goal is to estimate  $x_t$ given $\mathcal{Y}_t$. 
A natural criterion is the conditional minimum mean-square error (MMSE),
\begin{equation}\label{eqn:MMSE}
    \min_{\psi_t \in \mathcal{F}_t} 
    J_t(\psi_t, \Qdist_{e,t})
    \coloneqq 
    \mathbb{E}\left[\|x_t - \psi_t(\mathcal{Y}_t)\|^2 \mid \mathcal{Y}_{t-1}\right],
\end{equation}
where $\mathcal{F}_t$ denotes the set of measurable estimators $\psi_t:(\real{n_y})^{t+1}\to\real{n_x}$. 
That is, $\psi_t$ minimizes the conditional mean-square estimation error given past observations.

For $t>0$, the conditional distribution of $(x_t,y_t)$ given $\mathcal{Y}_{t-1}$ is fully
determined by the posterior distribution $\Postdistprev$ of $x_{t-1}$ and the joint noise
distribution $\Qdist_{e,t}\coloneqq\Qdist_{w,t-1}\otimes\Qdist_{v,t}$.
At $t=0$, no posterior from a previous step exists.
We adopt the convention $\mathcal{Y}_{-1}=\emptyset$ and define
$\Pdist_{x,0}^-\coloneqq\Qdist_{x,0}$ and
$\Qdist_{e,0}\coloneqq\Pdist_{x,0}^-\otimes\Qdist_{v,0}$, so that
\eqref{eqn:MMSE} also covers the initial stage.\footnote{We use the superscript ``$-$'' for prior quantities; e.g., $\xprior_t$, $\Xprior$, and $\Pdist_{x,t}^{-}$ denote the prior mean, covariance, and distribution of $x_t$ conditioned on $\mathcal{Y}_{t-1}$.}

When $x_0 \sim \Gauss(\hat{x}_0^-, \Xpriorinit)$ and $\{w_t\}, \{v_t\}$ are independent Gaussian sequences with known covariances, the optimal causal MMSE estimator reduces to the classical KF.
In practice, however, noise statistics are typically estimated from limited data and are thus
subject to ambiguity.
 In practice, only nominal distributions $\hat{\Qdist}_{w,t}$, $\hat{\Qdist}_{v,t}$, and $\hat{\Qdist}_{x,0}$ are available from identification procedures~(e.g.,~\cite{mehra1970identification,odelson2006new}).\footnote{A hat, $\hat{\cdot}$, denotes nominal quantities; e.g., $\hat{x}_0^-$ and $\hat{\Qdist}_{x,0}$ represent the nominal initial state mean and distribution, respectively.} 
Even modest deviations between nominal and true distributions can significantly degrade performance or cause divergence, especially when disturbances accumulate over time (e.g., marginally stable or unstable systems).

To address this, we formulate a stage-wise minimax problem in which the estimator competes against an adversary that selects distributions from ambiguity sets $\ambset_{j,t} \subset \mathcal{P}(\real{n_j})$ for $j \in \{w,v,x\}$ to be defined later. 
The resulting DRSE problem at time $t$ is 
\begin{equation}\label{eqn:DRSE}
    \min_{\psi_t \in \mathcal{F}_t} \max_{\Pdist_{e,t} \in \ambset_{e,t}}
    J_t(\psi_t, \Pdist_{e,t}),
\end{equation}
where $\ambset_{e,0} \coloneqq \ambset_{x,0} \times \ambset_{v,0}$ and $\ambset_{e,t} \coloneqq \ambset_{w,t-1} \times \ambset_{v,t}$ for $t>0$. 
The corresponding joint distributions are  $\Pdist_{e,0} \coloneqq \Priordistinit \otimes \Pdist_{v,0}$ and $\Pdist_{e,t} \coloneqq \Pdist_{w,t-1} \otimes \Pdist_{v,t}$, reflecting the assumed stage-wise independence
between process and measurement noise.

\begin{remark}\label{remark1}
In contrast to formulations that place ambiguity on a joint prior state--measurement distribution
(e.g.,~\cite{nguyen2023bridging,jang2025steady}), our approach imposes ambiguity directly on the
noise distributions.
This enforces a form of \emph{dynamic consistency}: admissible priors arise only through the
system dynamics~\eqref{eqn:system_eq} under admissible noise realizations. 
By contrast, previous DRSE formulations may allow priors that are not dynamically reachable. 
Our formulation also allows the adversary to couple estimation performance across time, capturing long-term accumulation of disturbances via the state recursion, even though the ambiguity sets themselves are stage-wise. 
For systems with slow or unstable dynamics, even small biases in $w_t$ can lead to unbounded state growth, making this sequential robustness essential.
\end{remark}

\subsection{Wasserstein Ambiguity Sets} \label{sec:Wass_and_Gelbr}

 We adopt Wasserstein ambiguity sets due to their favorable analytical properties and their
ability to exclude statistically irrelevant distributions (e.g.,~\cite{esfahani2015data,gao2023distributionally}).
The \emph{type-2 Wasserstein distance} between two probability distributions $\Pdist,\Qdist \in \mathcal{P}(\mathcal{W})$ is defined as
\[
    \Wass{\Pdist}{\Qdist}
    \coloneqq
    \inf_{\tau \in \mathcal{T}(\Pdist,\Qdist)}
    \Bigg\{
    \left(\int_{\mathcal{W} \times \mathcal{W}} \|x-y\|^2\, d\tau(x,y)\right)^{\frac{1}{2}}
    \Bigg\},
\]
where $\mathcal{T}(\Pdist,\Qdist)$ denotes the set of joint probability measures  with marginals $\Pdist$ and $\Qdist$. 
The optimization variable $\tau$ represents the \emph{transportation plan} that redistributes probability mass from $\Pdist$ to $\Qdist$.\footnote{We use the Euclidean norm to measure the transportation cost $\|x-y\|$.}

We define the ambiguity sets as Wasserstein balls centered at nominal distributions: for $j\in\{w, v, x_0\}$,
\begin{equation*}
\ambset_{j,t} \coloneqq \big\{ \Pdist_{j,t} \in \mathcal{P}(\real{n_j}) \,\vert \, \Wass {\Pdist_{j,t}}{\Qhat_{j,t}} \leq \theta_{j} \big\}, 
\end{equation*}
 where $\theta_j$ specifies the robustness radius.\footnote{The formulation can be extended to
correlated noise by placing a single Wasserstein ball around the joint distribution, at the cost
of higher-dimensional optimization problems.}

In general, computing $W_2$ is difficult. A useful surrogate is the \emph{Gelbrich distance}. Specifically, 
    the Gelbrich distance between $\Pdist \in \mathcal{P}(\real{n})$ and $\Qdist \in \mathcal{P}(\real{n})$ with mean vectors $\mu_1, \mu_2 \in \real{n}$ and covariance matrices $\Sigma_1, \Sigma_2\in \psd{n}$ is defined as
\begin{align*}
    \Gelb{\Pdist}{\Qdist} \coloneqq \sqrt{\|\mu_1-\mu_2\|_2^2 + \Bures^2(\Sigma_1, \Sigma_2)},
\end{align*}
where $\Bures(\Sigma_1, \Sigma_2) \coloneqq \sqrt{\Tr[\Sigma_1+\Sigma_2-2\big(\Sigma_2^{\frac{1}{2}} \Sigma_1 \Sigma_2^{\frac{1}{2}}\big)^{\frac{1}{2}}]}$ is the \emph{Bures--Wasserstein distance}.

The Gelbrich distance always provides a lower bound on the type-2 Wasserstein distance, i.e., $\Gelb{\Pdist}{\Qdist} \leq W_2(\Pdist,\Qdist)$. Moreover, equality holds when both distributions are Gaussian, or, more generally, elliptical with the same density generator~\cite{gelbrich1990formula}.

\section{Distributionally Robust Kalman Filters}\label{sec:drkf}

We develop tractable DR state estimators consistent with the noise-centric ambiguity model in~\Cref{sec:prelim}. 
We first present a finite-horizon formulation whose recursion mirrors that of the classical KF when driven by least-favorable noise covariances. 
We then specialize to the time-invariant case and derive a steady-state DRKF obtained from a \emph{single} stationary SDP, yielding a constant-gain filter with Kalman-level online complexity.

\subsection{Finite-Horizon Case}
We begin with a standard assumption on the nominal distributions that ensures tractability of the DRSE problem and preserves  the affine--Gaussian structure underlying closed-form KF updates.

\begin{assumption}\label{assump:Gauss}
    The nominal distributions of the initial state $x_0$, process noise $w_t$, and measurement noise $v_t$  are Gaussian for all $t$, i.e., $\Qhat_{x,0}=\Gauss(\xnom, \Xnom), \Qhat_{w,t}=\Gauss(\hat{w}_t,\hat{\Sigma}_{w,t})$, and $\Qhat_{v,t}=\Gauss(\hat{v}_t,\hat{\Sigma}_{v,t})$ with mean vectors $\xnom\in\real{n_x}, \hat{w}_t\in\real{n_x}$, and $\hat{v}_t\in\real{n_y}$ and covariance matrices $\Xnom \in \psd{n_x}, \hat{\Sigma}_{w,t}\in \psd{n_x}, \hat{\Sigma}_{v,t} \in \pd{n_y}$, respectively.
\end{assumption}

Under this assumption, the DRSE problem~\eqref{eqn:DRSE} admits the following convex reformulations.

\begin{lemma}
\label{lem:GelbrichMMSE} 
Suppose~\Cref{assump:Gauss} holds. Then, the DRSE problem~\eqref{eqn:DRSE} satisfies the following properties. 

$(i)$ At the initial stage, the following convex optimization problem has the same optimal value as~\eqref{eqn:DRSE}, and any of its optimal solutions are optimal for~\eqref{eqn:DRSE}:
\begin{equation} \label{eqn:DRMMSE_init_opt}
\begin{split}
\max_{\Xpriorinit,\Sigma_{v,0}}  & \Tr[\Xpriorinit -\Xpriorinit C_0^{\top}  (C_0\Xpriorinit C_0^{\top} + \Sigma_{v,0})^{-1} C_0 \Xpriorinit ] \\
\mbox{s.t.} \; & \Tr[\Xpriorinit + \Xnom -2 ((\Xnom)^{\frac{1}{2}} \Xpriorinit (\Xnom)^{\frac{1}{2}})^{\frac{1}{2}}] \leq \theta_{x_0}^2\\
& \Tr[\Sigma_{v,0} + \hat{\Sigma}_{v,0} -2 \big(\hat{\Sigma}_{v,0}^\frac{1}{2} \Sigma_{v,0} \hat{\Sigma}_{v,0}^\frac{1}{2}\big)^{\frac{1}{2}}] \leq \theta_{v}^2\\
& \Xpriorinit \succeq \lammin{\Xnom} I_{n_x}, \; \Sigma_{v,0} \succeq \lammin{\hat{\Sigma}_{v,0}} I_{n_y}\\
& \Xpriorinit\in\psd{n_x}, \; \Sigma_{v,0}\in\psd{n_y}.
\end{split}
\end{equation}

$(ii)$ For any $t>0$, fix the state distribution $\Postdistprev = \Gauss(\xpost_{t-1}, \Xpostprev)$. Then the following convex optimization problem has the same optimal value as~\eqref{eqn:DRSE}, and any of its optimal solutions are optimal for~\eqref{eqn:DRSE}:
\begin{equation}\label{eqn:DRMMSE_opt}
\begin{split}
\max_{\substack{\Sigma_{w,t-1},  \\\Sigma_{v,t}, \Xprior}} \; & \Tr[\Xprior-\Xprior C_t^{\top}  (C_t\Xprior C_t^{\top} + \Sigma_{v,t})^{-1} C_t \Xprior ] \\
\mbox{s.t.} \; & \Tr[\Sigma_{w,t-1} + \hat{\Sigma}_{w,t-1}  -2 \big(\hat{\Sigma}_{w,t-1}^\frac{1}{2} \Sigma_{w,t-1} \hat{\Sigma}_{w,t-1}^\frac{1}{2}\big)^{\frac{1}{2}}] \leq \theta_{w}^2\\
& \Tr[\Sigma_{v,t} + \hat{\Sigma}_{v,t} -2 \big(\hat{\Sigma}_{v,t}^\frac{1}{2} \Sigma_{v,t} \hat{\Sigma}_{v,t}^\frac{1}{2}\big)^{\frac{1}{2}}] \leq \theta_{v}^2\\
& \Xprior = A_{t-1} \Xpostprev A_{t-1}^\top + \Sigma_{w,t-1}\\
& \Sigma_{w,t-1} \succeq \lammin{\hat{\Sigma}_{w,t-1}} I_{n_x}, \; \Sigma_{v,t} \succeq \lammin{\hat{\Sigma}_{v,t}} I_{n_y}\\
&\Xprior, \Sigma_{w,t-1}\in\psd{n_x}, \;\Sigma_{v,t}\in\psd{n_y}.
\end{split}
\end{equation}

$(iii)$ If $(\Xpriorinitopt,\Sigma_{v,0}^*)$ and $(\Sigma_{w,t-1}^*,\Sigma_{v,t}^*)$ solve \eqref{eqn:DRMMSE_init_opt} and \eqref{eqn:DRMMSE_opt}, respectively, then the maximum in~\eqref{eqn:DRSE} is attained by the Gaussian distributions $\Priordistinitopt = \Gauss(\xnom, \Xpriorinitopt),  \Pdist_{w,t-1}^{*} = \Gauss(\hat{w}_{t-1}, \Sigma_{w,t-1}^{*})$ and $\Pdist_{v,t}^{*} = \Gauss(\hat{v}_t, \Sigma_{v,t}^{*})$ for each $t$. 
Furthermore, the minimax-optimal estimator is affine and given by
\begin{equation}\label{eqn:DR_estimator}
\begin{split}
\psi_{t}^*& (\mathcal{Y}_{t}) =  \xprior_t + \Xprior C_t^{\top} (C_t\Xprior  C_t^{\top} + \Sigma_{v,t}^{*})^{-1}(y_{t}-C_t\xprior_t -\hat{v}_{t}),
\end{split}
\end{equation}
with $\xprior_t = A_{t-1} \xpost_{t-1} + \hat{w}_{t-1}, \Xprior = A_{t-1} \Xpostprev A_{t-1}^\top + \Sigma_{w,t-1}^*$, and $\Xpriorinit = \Xpriorinitopt, \xprior_0 = \xnom$.
\end{lemma}

Problems~\eqref{eqn:DRMMSE_init_opt} and~\eqref{eqn:DRMMSE_opt} define the stage-wise minimax interaction: the estimator selects $\psi_t$, while the adversary selects noise covariances within Wasserstein balls. Problem~\eqref{eqn:DRMMSE_init_opt} yields the least-favorable prior and measurement covariances $(\Xpriorinitopt,\Sigma_{v,0}^{*})$ at $t=0$, whereas~\eqref{eqn:DRMMSE_opt} yields the least-favorable process and measurement covariances $(\Sigma_{w,t-1}^*, \Sigma_{v,t}^*)$ for $t \geq 1$. The corresponding optimal values represent the least-favorable posterior MSE at each stage.

\Cref{lem:GelbrichMMSE} extends~\cite[Thm.~3.1]{nguyen2023bridging} to the noise-centric setting by replacing ambiguity on the prior state (for $t>0$) with ambiguity on the process noise. 
 Although the Wasserstein constraints are equivalently expressible in terms of first and second moments,
this does not meaningfully restrict the adversary under~\Cref{assump:Gauss}. 
 For Gaussian nominal distributions, the least-favorable distributions are Gaussian, and the worst-case behavior is fully characterized by their means and covariances. Furthermore, even without Gaussian nominal assumptions,~\eqref{eqn:DR_estimator} remains minimax optimal within the class of affine estimators~\cite{nguyen2023bridging}.

\begin{remark}
Although the ambiguity sets are defined as $W_2$ balls over full distributions, the maximization problems in~\eqref{eqn:DRMMSE_init_opt}
and~\eqref{eqn:DRMMSE_opt} admit least-favorable distributions that preserve the nominal noise means. Indeed, for any fixed distribution, the minimizing
estimator is the conditional mean, and the  conditional MSE depends only on the posterior covariance through its trace. Mean shifts consume Wasserstein budget without increasing the worst-case MSE as effectively as perturbing the covariance. Hence, mean shifts are suboptimal for the adversary, and the ambiguity budget is
used entirely to perturb covariances, as reflected in \Cref{lem:GelbrichMMSE} and~\Cref{thm:DRKF}.
\end{remark}

\begin{remark}\label{rem:redundant_bounds}
By adapting the structural argument used in~\cite[Lemma~A.3]{nguyen2023bridging}, the DRSE problem~\eqref{eqn:DRSE} admits at least one maximizer whose covariance matrices satisfy $\Xpriorinitopt \succeq \lammin{\Xnom} I_{n_x}, \Sigma_{v,0}^* \succeq \lammin{\hat{\Sigma}_{v,0}} I_{n_y}$,
for $t=0$, and $\Sigma_{w,t-1}^* \succeq \lammin{\hat{\Sigma}_{w,t-1}} I_{n_x},\Sigma_{v,t}^* \succeq \lammin{\hat{\Sigma}_{v,t}} I_{n_y}$ for all $t > 0$ (see Appendix~\ref{app:ev_lower}). Because such a maximizer always exists, the explicit eigenvalue lower-bound constraints in~\eqref{eqn:DRMMSE_init_opt} and~\eqref{eqn:DRMMSE_opt} are \emph{redundant}: they do not alter the optimal value but only restrict attention to maximizers that satisfy these bounds. Equivalently, removing these constraints yields problems with the same
optimal value as~\eqref{eqn:DRSE}.
\end{remark}

Both problems~\eqref{eqn:DRMMSE_init_opt} and~\eqref{eqn:DRMMSE_opt} are solvable for $\hat{\Sigma}_{v,t} \in \pd{n_y}$ since their objectives are continuous over compact feasible sets. Applying the standard Schur complement argument,
these problems reduce to SDPs that can be solved efficiently using off-the-shelf solvers.
\begin{corollary}\label{cor:sdp}
    Suppose~\Cref{assump:Gauss} holds. Then, for any time $t>0$, the optimization problem~\eqref{eqn:DRMMSE_opt} is equivalent to the following tractable convex SDP:
    \begin{equation} \label{eqn:DRKF_SDP}
\begin{split}
\max_{\substack{\Xprior, \Xpost, \Sigma_{w,t-1},\\ \Sigma_{v,t},Y,Z}} \; & \Tr[\Xpost ] \\
\mbox{s.t.} \; & \begin{bmatrix} \Xprior - \Xpost & \Xprior C_t^{\top} \\ C_t \Xprior & C_t \Xprior C_t^{\top} + \Sigma_{v,t} \end{bmatrix} \succeq 0\\
& \begin{bmatrix} \hat{\Sigma}_{w,t-1} & Y \\ Y^{\top} & \Sigma_{w,t-1} \end{bmatrix} \succeq 0, \; \begin{bmatrix} \hat{\Sigma}_{v,t} & Z \\ Z^{\top} & \Sigma_{v,t} \end{bmatrix} \succeq 0 \\
&\Tr[\Sigma_{w,t-1} + \hat{\Sigma}_{w,t-1}  -2 Y]  \leq \theta_w^2 \\
&\Tr[\Sigma_{v,t} + \hat{\Sigma}_{v,t} -2 Z] \leq \theta_v^2\\
& \Xprior = A_{t-1} \Xpostprev A_{t-1}^\top + \Sigma_{w,t-1}\\
& \Sigma_{w,t-1} \succeq \lammin{\hat{\Sigma}_{w,t-1}} I_{n_x}, \; \Sigma_{v,t} \succeq \lammin{\hat{\Sigma}_{v,t}} I_{n_y}\\
& \Xprior, \Xpost, \Sigma_{w,t-1} \in \psd{n_x}, \; \Sigma_{v,t} \in \psd{n_y}\\
&Y\in\real{n_x\times n_x},\; Z \in\real{n_y \times n_y}.
\end{split}
\end{equation}
\end{corollary}

Similar results hold for  $t=0$. Solving \eqref{eqn:DRKF_SDP}  offline for all $t>0$ yields the least-favorable covariances $(\Sigma_{w,t-1}^*,\Sigma_{v,t}^*)$.
The next theorem shows that these covariances can be used directly in a  Kalman-style recursion, resulting in a DR state estimator that is robust to distributional uncertainty.

\begin{theorem}[DR Kalman Filter]\label{thm:DRKF}
    Under~\Cref{assump:Gauss}, the minimax-optimal DR state estimate $\psi^*_t(\mathcal{Y}_t)$ at any time $t$ coincides with the conditional mean $\xpost_t$ under the least-favorable distributions. Moreover, the least-favorable prior and posterior remain Gaussian: $\Priordistopt=\Gauss(\xprior_t,\Xprior)$, $\Postdistopt=\Gauss(\xpost_t,\Xpost)$. These distributions are recursively computed for $t=0,1, \dots$ as follows:
\begin{itemize}
\item \textbf{(Measurement Update)} Compute the DR Kalman gain 
\begin{equation}\label{eqn:Kalman_gain}
K_t = \Xprior C_t^{\top} (C_t { \Xprior } C_t^{\top} + \Sigma_{v,t}^{*})^{-1},
\end{equation}
where $\Sigma_{v,t}^{*}$ is the maximizer of~\eqref{eqn:DRMMSE_opt} for $t > 0$ and~\eqref{eqn:DRMMSE_init_opt} for $t=0$.
Then, update $\xpost_t$ and $\Xpost$  as
\begin{align}
&\xpost_t = K_t (y_{t}-C_t \xprior_t -\hat{v}_{t}) + \xprior_t \label{eqn:cond_mean}\\
&\Xpost = (I - K_t  C_t) \Xprior.\label{eqn:cond_cov}
\end{align}
The initial values are $\xprior_0 = \xnom, \Xpriorinit = \Xpriorinitopt$, where $\Xpriorinitopt$ is the maximizer of~\eqref{eqn:DRMMSE_init_opt} at $t=0$.

\item \textbf{(State Prediction)} Predict $\xprior_{t+1}$ and $\Xpriornext$ as
\begin{align}
    \begin{split}
        \xprior_{t+1} &= A_t\xpost_{t} + \hat{w}_t \label{eqn:cond_mean_prior}
    \end{split} \\
    \Xpriornext &= A_t \Xpost A_t^\top + \Sigma_{w,t}^{*},\label{eqn:cond_cov_prior}
\end{align}
where $\Sigma_{w,t}^{*}$ is the maximizer of~\eqref{eqn:DRMMSE_opt} at time $t+1$.
\end{itemize}
\end{theorem}

Notably, the measurement update and state prediction equations~\eqref{eqn:cond_mean}--\eqref{eqn:cond_cov_prior} mirror those of the classical KF when the system is driven by the least-favorable distributions.

\subsection{Infinite-Horizon Case}\label{subsec:infDRKF}

While the DRKF in~\Cref{thm:DRKF} provides a tractable solution to the finite-horizon DRSE problem, our primary interest lies in its long-term behavior. To study this regime, we assume that both the system dynamics and the nominal uncertainty statistics are time-invariant.

\begin{assumption}\label{assump:time_inv_nom}
    The system~\eqref{eqn:system_eq} is time-invariant, i.e., $A_t \equiv A, C_t \equiv C$. Moreover, the nominal uncertainty distributions are stationary, so that $\hat{w}_t \equiv \hat{w}, \hat{v}_t \equiv \hat{v}$ and $\hat{\Sigma}_{v,t} \equiv \hat{\Sigma}_v, \hat{\Sigma}_{w,t} \equiv\hat{\Sigma}_w$ for all $t \geq 0$, where $\hat{\Sigma}_w \in \pd{n_x}$ and $\hat{\Sigma}_v \in \pd{n_y}$. The nominal initial state covariance satisfies $\Xnom\in\pd{n_x}$.
\end{assumption}

Under~\Cref{assump:time_inv_nom}, the stage-wise SDPs~\eqref{eqn:DRMMSE_opt}--\eqref{eqn:DRKF_SDP} reduce to a time-invariant recursion for the covariance matrices. The resulting sequences of least-favorable noise covariances $\{\Sigma_{w,t}^*, \Sigma_{v,t}^*\}$ and state covariances $\{\Xprior,\Xpost\}$ may converge to limiting values $\{\Sigma_{w,\infty}, \Sigma_{v,\infty}\}$ and $\{\Xpriorss, \Xpostss\}$ as $t \to \infty$. Whether such convergence holds depends on additional conditions, which are established later in~\Cref{subsec:convergence}. Motivated by this, we seek a steady-state DR estimator with a constant gain.  

In the infinite-horizon setting, the DRSE problem reduces to computing these steady-state covariances and the associated DR Kalman gain. Since the dynamics and nominal statistics are time-invariant, the stage-wise SDP~\eqref{eqn:DRMMSE_opt} simplifies to the following   convex program:
\begin{equation*}\label{eqn:DRMMSE_opt_ss}
\begin{split}
\max_{\substack{\Sigma_{w,\infty}, \Sigma_{v,\infty} \\ \Xpriorss, \Xpostss}} \; & \Tr[\Xpostss ] \\
\mbox{s.t.} \; & \Xpostss = \Xpriorss  -\Xpriorss C^{\top}  (C\Xpriorss C^{\top} + \Sigma_{v,\infty})^{-1} C \Xpriorss\\
& \Xpriorss = A \Xpostss A^\top + \Sigma_{w,\infty}\\
& \Tr[\Sigma_{w,\infty} + \hat{\Sigma}_{w} -2 \big(\hat{\Sigma}_{w}^\frac{1}{2} \Sigma_{w,\infty} \hat{\Sigma}_{w}^\frac{1}{2}\big)^{\frac{1}{2}}] \leq \theta_{w}^2\\
&\Tr[\Sigma_{v,\infty} + \hat{\Sigma}_{v} -2 \big(\hat{\Sigma}_{v}^\frac{1}{2} \Sigma_{v,\infty} \hat{\Sigma}_{v}^\frac{1}{2}\big)^{\frac{1}{2}}] \leq \theta_{v}^2\\
& \Sigma_{w,\infty} \succeq \lammin{\hat{\Sigma}_{w}} I_{n_x}, \; \Sigma_{v,\infty} \succeq \lammin{\hat{\Sigma}_{v}} I_{n_y}\\
&\Xpriorss, \Xpostss, \Sigma_{w,\infty}\in\psd{n_x},\Sigma_{v,\infty}\in\psd{n_y}.
\end{split}
\end{equation*}
As in the finite-horizon case, this problem admits an equivalent SDP formulation.
In particular, it can be written as the following \emph{single} stationary SDP:
\begin{equation} \label{eqn:DRKF_SDP_ss}
\begin{split}
\max_{\substack{\Xpriorss ,\Xpostss,\\ \Sigma_{w,\infty}, \Sigma_{v,\infty},\\ Y,Z}} \; & \Tr[\Xpostss ] \\
\mbox{s.t.} \; & \begin{bmatrix} \Xpriorss - \Xpostss & \Xpriorss C^{\top} \\ C \Xpriorss & C \Xpriorss C^{\top} + \Sigma_{v,\infty} \end{bmatrix} \succeq 0\\
& \begin{bmatrix} \hat{\Sigma}_{w} & Y \\ Y^{\top} & \Sigma_{w,\infty} \end{bmatrix} \succeq 0, \quad \begin{bmatrix} \hat{\Sigma}_{v} & Z \\ Z^{\top} & \Sigma_{v,\infty} \end{bmatrix} \succeq 0 \\
& \Xpriorss = A \Xpostss A^{\top} + {\Sigma}_{w,\infty} \\
&\Tr[\Sigma_{w, \infty} + \hat{\Sigma}_{w}  -2 Y]  \leq \theta_w^2 \\
&\Tr[\Sigma_{v,\infty} + \hat{\Sigma}_{v} -2 Z] \leq \theta_v^2\\
& \Sigma_{w,\infty} \succeq \lammin{\hat{\Sigma}_{w}} I_{n_x}, \; \Sigma_{v,\infty} \succeq \lammin{\hat{\Sigma}_{v}} I_{n_y}\\
& \Xpriorss, \Xpostss, \Sigma_{w,\infty} \in \psd{n_x}, \; \Sigma_{v,\infty} \in \psd{n_y}\\
&Y\in\real{n_x\times n_x},\; Z \in\real{n_y \times n_y}.
\end{split}
\end{equation}
The optimal solution $\{\Sigma_{w,\infty}^*, \Sigma_{v,\infty}^*, \Xpriorssopt, \Xpostssopt\}$ defines the least-favorable stationary covariances and yields the steady-state DR Kalman gain
\begin{equation}\label{eqn:ss_gain}
K_{\infty}^* := \Xpriorssopt C^{\top} (C\Xpriorssopt  C^{\top}+ \Sigma_{v,\infty}^*)^{-1}.
\end{equation}
The resulting time-invariant  estimator is
\begin{equation} \label{eqn:DR_est_update} 
\xpost_t = \psi_{\infty}^*(\mathcal{Y}_{t}) := \xprior_t +  K_{\infty}^* (y_{t}-C\xprior_t -\hat{v}), 
\end{equation}
with the prior update
\begin{equation}\label{eqn:cond_mean_prior_ss}
\xprior_{t+1} = A \xpost_t + \hat{w}.
\end{equation}

Compared with the finite-horizon DRKF, the steady-state filter~\eqref{eqn:DR_est_update} requires solving only a single stationary SDP offline.
This makes it particularly suitable for real-time implementation. More importantly,  it directly optimizes  worst-case asymptotic performance, providing robustness to persistent disturbances and long-term distributional errors--an essential property for safety-critical systems with slow or unstable dynamics.

\begin{algorithm}[t]
\caption{Time-Varying DRKF}
\label{alg:DRKF}
\begin{algorithmic}[1]
\Require System matrices $\{A_t, C_t\}_{t=0}^{T-1}$, 
nominal means and covariances $(\hat{w}_t, \hat{v}_t, \hat{\Sigma}_{w,t}, \hat{\Sigma}_{v,t})$, 
prior state mean $\xnom$ and covariance $\Xnom$,
ambiguity radii $(\theta_w, \theta_v, \theta_{x_0})$, 
horizon length $T$

\State Solve~\eqref{eqn:DRMMSE_init_opt} to obtain $(\Xpriorinitopt, \Sigma_{v,0}^*)$\Comment{Offline stage}
\State Compute DR gain $K_0$ via~\eqref{eqn:Kalman_gain} and update the posterior covariance $\Xpostinit$ 
via~\eqref{eqn:cond_cov}
\For{$t = 1, \dots, T-1$}
    \State Solve~\eqref{eqn:DRKF_SDP} to obtain $(\Sigma_{w,t-1}^*, \Sigma_{v,t}^*,\Xprior,\Xpost)$
    \State Compute DR gain $K_t$ via~\eqref{eqn:Kalman_gain}
\EndFor
\State Set $\xprior_0 = \xnom$ \Comment{Online stage}
\For{$t = 0,\dots,T-1$}  
    \State Observe $y_t$
    \State Update the state estimate $\xpost_t$ via~\eqref{eqn:cond_mean} using $K_t$
    \State Predict the prior mean $\xprior_{t+1}$ via~\eqref{eqn:cond_mean_prior}
\EndFor
\end{algorithmic}
\end{algorithm}

\begin{algorithm}[t]
\caption{Steady-State DRKF}
\label{alg:SSDRKF}
\begin{algorithmic}[1]
\Require System matrices $(A, C)$,
nominal means and covariances $(\hat{w}, \hat{v}, \hat{\Sigma}_w, \hat{\Sigma}_v)$,
prior state mean $\xnom$, ambiguity radii $(\theta_w, \theta_v)$

\Statex \Comment{Offline stage}
\State Solve~\eqref{eqn:DRKF_SDP_ss} to obtain $(\Sigma_{w,\infty}^*, \Sigma_{v,\infty}^*, \Xpriorssopt, \Xpostssopt)$ 
\State Compute the steady-state DR gain via~\eqref{eqn:ss_gain}
\State Set $\xprior_0 = \xnom$ \Comment{Online stage}
\For{$t = 0,1,2,\dots$} 
    \State Observe $y_t$
    \State Update the state estimate $\xpost_t$ via~\eqref{eqn:DR_est_update} using $K_\infty^*$
    \State Predict the prior mean $\xprior_{t+1}$ via~\eqref{eqn:cond_mean_prior_ss}
\EndFor
\end{algorithmic}
\end{algorithm}

\subsection{Algorithms and Complexity}

We summarize the proposed DRKF methods for both the time-varying and steady-state settings.
The two variants differ only in how the noise covariances (and hence the Kalman gains) are
computed offline to ensure robustness against distributional ambiguity.

\subsubsection{Time-Varying and Steady-State DRKF}

As shown in~\Cref{alg:DRKF}, the time-varying DRKF is implemented by solving the stage-wise SDP problems~\eqref{eqn:DRMMSE_init_opt} and~\eqref{eqn:DRMMSE_opt} offline over a horizon $T$.
This yields the least-favorable noise covariances and the corresponding DR gains $\{K_t\}$.
If the nominal covariances are known \emph{a~priori}, these gains can be fully precomputed.

Under the time-invariant assumptions of~\Cref{subsec:infDRKF}, the DRKF recursion may converge to a
stationary solution.
In this case, the offline stage reduces to solving the \emph{single} stationary SDP
\eqref{eqn:DRKF_SDP_ss}, which yields the least-favorable noise covariances, the steady-state
prior and posterior covariances, and the constant DR Kalman gain $K_\infty^*$.

In both cases, the online stage performs only the standard KF measurement update and prediction,
and thus has the same per-step computational complexity as the classical KF.

\subsubsection{Discussion on Computational Complexity}

The computational cost of the proposed DRKF methods is dominated by the \emph{offline} SDP
computations; the \emph{online} recursion has the same per-step complexity as the classical KF.

In the time-varying case, each stage requires solving the SDP~\eqref{eqn:DRKF_SDP}, whose decision variables include covariance blocks of size $n_x \times n_x$ and $n_y \times n_y$. 
This yields $\mathcal{O}(n_x^2+n_y^2)$ free variables, with the largest linear matrix inequality (LMI) block of dimension $\mathcal{O}(n_x+n_y)$. 
Using  standard interior-point methods~\cite{nemirovski2004interior}, the per-stage complexity is $\tilde{\mathcal{O}}\left((n_x+n_y)^{3.5}\right)$,\footnote{Here,  $\tilde{\mathcal{O}}(\cdot)$  suppresses polylogarithmic factors.} 
and the total offline cost over a horizon $T$ scales as
$\tilde{\mathcal{O}}\left((n_x+n_y)^{3.5}T\right)$.

For comparison, the stage-wise DRKF of~\cite{NEURIPS_DRKF} and the DR-MMSE estimator of~\cite{nguyen2023bridging} also require solving a convex program at each stage. 
The former relies on joint state--measurement covariances of dimension $(n_x+n_y)$, leading to
similar polynomial scaling but with larger LMI blocks and constants.
Like our time-varying DRKF, these methods scale linearly with $T$, but they do
not admit a steady-state formulation that reduces the offline computation to a \emph{single} SDP.

Temporally coupled ambiguity sets, as in~\cite{kargin2024distributionally}, model long-range dependence across the noise sequence at the cost of significantly higher complexity: all $T$ stages are merged into a single horizon-level SDP whose size grows linearly with $T$, resulting in interior-point complexity of  $\tilde{\mathcal{O}}(n_x^6 T^6)$.
By contrast, our noise-centric formulation solves a \emph{fixed-size} SDP at each stage, so the offline cost grows only linearly in $T$.

In the infinite-horizon setting,~\cite{kargin2024distributionally} characterizes the steady-state filter via a frequency-domain max-min problem over Toeplitz operators, producing a generally non-rational filter that requires spectral factorization and approximation for implementation. No finite-dimensional formulation is available. In contrast, our steady-state DRKF is obtained from a \emph{single} stationary SDP of size $(n_x+n_y)$ with offline complexity $\tilde{\mathcal{O}}\left((n_x+n_y)^{3.5}\right)$. It directly yields a constant gain $K_\infty^*$ and an online recursion matching the classical steady-state KF.
Although this stage-wise ambiguity does not model temporal noise correlations, it preserves the separability of the Riccati recursion and keeps the DR filter tractable for real-time use.

\section{Theoretical Analysis}\label{sec:theoretical}

Having established the DRKF formulations in both the finite- and infinite-horizon settings, we now analyze their fundamental theoretical properties. Our goal is to show that the DRKF retains key structural guarantees of the classical KF, such as stability and boundedness, while providing provable robustness under distributional uncertainty. In particular, we demonstrate that the DRKF covariance recursion remains spectrally bounded and that its steady-state solution inherits standard convergence properties of the classical filter.

\subsection{Spectral Boundedness}\label{subsec:spectral_bound}

We begin by quantifying how Wasserstein ambiguity sets constrain the least-favorable covariance matrices, and how these constraints propagate through the DRKF recursion in~\Cref{thm:DRKF}. 
This analysis shows that the DRKF operates within a uniformly bounded spectral envelope, a property that underpins its stability and distinguishes it from several existing robust formulations.

Our starting point is a well-known characterization of the (non-squared) Bures--Wasserstein distance.

\begin{lemma}\label{lem:procr} \cite[Thm.~1]{bhatia2019bures}
Let $A,B \in \psd{n}$. The Bures--Wasserstein distance satisfies
\begin{equation*}\label{eqn:Procrustes}
    \Bures(A,B) \;=\; \min_{U \in O(n)} \| A^{1/2} - B^{1/2}U \|_{F},
\end{equation*}
where $O(n)$ denotes the orthogonal group. The minimizer exists and is given by the orthogonal factor in the polar decomposition of $B^{1/2}A^{1/2}$.
\end{lemma}

This characterization enables sharp spectral bounds for matrices within a Bures--Wasserstein ball.

\begin{proposition}\label{prop:bures_bound}
Fix $\hat{X} \in\psd{n}$. For any $X \in\psd{n}$ satisfying $\Bures(X,\hat{X}) \leq \theta$, the minimum and maximum eigenvalues of $X$ are bounded as
\begin{equation*}
    \underline{\lambda}(\hat{X},\theta) \;\leq\; \lammin{X} 
    \;\leq\; \lammax{X} \;\leq\; \overline{\lambda}(\hat{X},\theta),
\end{equation*}
where
\begin{align*}
    \underline{\lambda}(\hat{X},\theta) &\coloneqq \left(\max \left \{0,\sqrt{\lammin{\hat{X}}}-\theta \right\}\right)^2,\\
    \overline{\lambda}(\hat{X},\theta) &\coloneqq \left(\sqrt{\lammax{\hat{X}}}+\theta\right)^2.
\end{align*}
\end{proposition}

The proof is provided in Appendix~\ref{app:bures_bound}.
This result shows that a Bures--Wasserstein ball induces both upper and lower
spectral bounds on feasible covariance matrices, defining an \emph{eigenvalue
tube} around the nominal covariance.
This geometric perspective is essential: it rules out pathological behavior in
which worst-case covariances can become arbitrarily inflated or collapse in
directions that are inconsistent with the nominal statistics.
As a result, the DRKF operates within a controlled spectral envelope under
distributional perturbations allowed by the ambiguity sets.
By contrast, KL-based ambiguity sets may admit least-favorable distributions
with highly distorted or ill-conditioned covariance structure, leading to
potentially pathological estimators~\cite{gao2023distributionally}.

We now translate these spectral bounds to the least-favorable noise covariances that enter the DRKF recursion.

\begin{corollary}\label{cor:bounds}    
Let $\hat{\Sigma}_{v,t}\in\pd{n_y}$ and $\hat{\Sigma}_{w,t}\in\psd{n_x}$ denote the nominal noise covariances. For ambiguity radii $\theta_v,\theta_w, \theta_{x_0}$, define 
\begin{equation}\label{eqn:noisebound}
\begin{split}
    \underline{\lambda}_{v,t} &:= \lammin{\hat{\Sigma}_{v,t}},\qquad
    \overline{\lambda}_{v,t} := \overline{\lambda}(\hat{\Sigma}_{v,t},\theta_v),\\
    \underline{\lambda}_{w,t} &:= \lammin{\hat{\Sigma}_{w,t}},\qquad
    \overline{\lambda}_{w,t} := \overline{\lambda}(\hat{\Sigma}_{w,t},\theta_w),\\
     \underline{\lambda}_{x,0} &:= \lammin{\Xnom},\qquad
    \overline{\lambda}_{x,0} := \overline{\lambda}(\Xnom,\theta_{x_0}).   
\end{split}
\end{equation}
Then,  the least-favorable covariances $\Sigma_{v,t}^*, \Sigma_{w,t}^*$ and $\Xpriorinitopt$ obtained from~\eqref{eqn:DRMMSE_init_opt}--\eqref{eqn:DRMMSE_opt} satisfy
\begin{align*}
    &\underline{\lambda}_{v,t} I_{n_y} \preceq \Sigma_{v,t}^* \preceq \overline{\lambda}_{v,t} I_{n_y},\\
    &\underline{\lambda}_{w,t} I_{n_x} \preceq \Sigma_{w,t}^* \preceq \overline{\lambda}_{w,t} I_{n_x}, \\
    &\underline{\lambda}_{x,0} I_{n_x} \preceq \Xpriorinitopt \preceq \overline{\lambda}_{x,0} I_{n_x}.
\end{align*}
\end{corollary}

The proof is provided in Appendix~\ref{app:bounds}.
In the time-invariant setting, the constants
$\underline{\lambda}_w \coloneq \lammin{\hat{\Sigma}_w}$, $\underline{\lambda}_v \coloneq \lammin{\hat{\Sigma}_v}$, $\overline{\lambda}_w \coloneq \overline{\lambda}(\hat{\Sigma}_w,\theta_w)$, and
$\overline{\lambda}_v \coloneq \overline{\lambda}(\hat{\Sigma}_v,\theta_v)$ are fixed. 
Thus,~\Cref{cor:bounds} ensures that the DRKF operates within a compact spectral envelope determined solely by the nominal statistics and the Wasserstein radii.

\begin{theorem}\label{thm:drkf_sandwich}
Consider two classical KFs with noise covariances chosen according to the lower and upper bounds in~\eqref{eqn:noisebound}. Let $(\Xpriorlow,\Xpostlow)$ and $(\Xpriorhigh,\Xposthigh)$ denote their respective prior and posterior error covariances, with initial conditions $\Xpriorlowinit = \underline{\lambda}_{x,0} I_{n_x}$ and $\Xpriorhighinit = \overline{\lambda}_{x,0} I_{n_x}$. Then, for all $t\geq 0$, the DRKF covariances $(\Xprior,\Xpost)$ satisfy 
\begin{equation*}\label{eqn:drkf_cov_sandwich}
\Xpriorlow \preceq \Xprior \preceq \Xpriorhigh,\qquad
\Xpostlow \preceq \Xpost \preceq \Xposthigh.
\end{equation*}
\end{theorem}

The proof is provided in Appendix~\ref{app:drkf_sandwich}.
Given the initial prior state covariance, the lower and upper bounds in~\Cref{thm:drkf_sandwich} can be precomputed by running two standard KF recursions before solving the optimization problems~\eqref{eqn:DRMMSE_init_opt} (for $t=0$) and~\eqref{eqn:DRMMSE_opt} (for $t>0$).

This spectral boundedness is a direct consequence of our ambiguity set design, which models uncertainty separately for the process and measurement noise distributions. In contrast, previous DRKFs based on ambiguity sets centered on the joint prior-measurement distributions, as in \cite{jang2025steady}, do not directly admit such precomputed spectral bounds. This highlights an additional advantage of our noise-centric construction, complementing the benefit noted in~\Cref{remark1}.

\subsection{Convergence}\label{subsec:convergence}

Having established the key theoretical properties of the least-favorable covariance matrices, we now derive conditions under which the DRKF admits a unique steady-state solution, and the sequence of time-varying gains converges to its steady-state value. Specifically, recall that the stationary DR Kalman gain $K_\infty^*$ is defined in~\eqref{eqn:ss_gain}. Our goal is to show that 
\[
\lim_{t\to\infty} K_t = K_\infty^*.
\]
Under the standing assumptions, the time-varying SDP problems
\eqref{eqn:DRMMSE_init_opt} and \eqref{eqn:DRMMSE_opt} admit optimal solutions (the feasible sets are nonempty and compact under the Bures--Wasserstein bounds), and
therefore the sequence of DR Kalman gains $\{K_t\}$ is well defined.
In the
time-invariant setting, the steady-state SDP~\eqref{eqn:DRKF_SDP_ss} is feasible
(e.g., by choosing the nominal covariances), and below we show that the DR Riccati
recursion converges to a unique fixed point characterized by this SDP.

The prior covariance recursion~\eqref{eqn:cond_cov_prior} can be equivalently expressed as
\begin{equation}\label{eqn:riccati_like}
    \Xpriornext = A\bigl((\Xprior)^{-1} + C^\top (\Sigma_{v,t}^*)^{-1} C\bigr)^{-1} A^\top + \Sigma_{w,t}^*,
\end{equation}
with $\Xpriorinit = \Xpriorinitopt$.
If $\Xnom \succ 0$, then $\Xpriorinit \succ 0$ and the recursion is well defined.\footnote{
If $\Xnom \succeq 0$, the uniform bound $\Sigma_{w,t}^* \succeq \underline{\lambda}_w I$
ensures that $\Sigma_{x,1}^- \in \pd{n_x}$, so the information-form recursion is well
defined from time $t=1$ onward.}
This motivates the following definition of the \emph{DR Riccati mapping}:
\begin{equation}\label{eqn:DR_Riccati_def}
     r_{\ambset}(\Xprior) := A\bigl((\Xprior)^{-1} + C^\top (\Sigma_{v,t}^*)^{-1} C\bigr)^{-1} A^\top + \Sigma_{w,t}^*.
\end{equation}
Under~\Cref{assump:time_inv_nom}, the problem data of the stage-wise SDP are time-invariant. Accordingly, the sequence $\{(\Sigma_{w,t}^*,\Sigma_{v,t}^*) \}$ is generated by repeatedly applying the same SDP template along the state covariance trajectory.
Observing the structural similarity with Riccati equations arising in robust and risk-sensitive filtering (e.g.,~\cite{whittle1981risk,jacobson1973optimal}), we pursue contraction-based analysis in this subsection.

We begin by imposing the standard observability assumption.\footnote{Since $\hat{\Sigma}_w \succ 0$ and $\Sigma_{w,t}^* \succeq \underline{\lambda}_w I$, the process noise covariance is uniformly positive definite; hence no additional controllability assumption is required.}
\begin{assumption}\label{assump:obsv}
    The pair $(A,C)$ is observable.
\end{assumption}

To establish strict contraction under the observability assumption, we employ a
downsampled (lifted) representation of the DR Riccati recursion, since the
one-step mapping may fail to be contractive due to rank-deficient measurement
operators, whereas observability guarantees full column rank of the lifted
observability matrix.

Fix $N\in\mathbb{N}$ and consider the system~\eqref{eqn:system_eq} under the least-favorable distributions. Define the downsampled state as $x_t^d \coloneq x_{tN}$. Downsampling yields an $N$-block lifted model with block-diagonal uncertainty structure, enabling a contraction argument similar to~\cite{zorzi2015convergence,zorzi2017convergence}. Let $\bm{w}_t^N \coloneqq [w_{tN+N-1}^\top, \dots, w_{tN}^\top]^\top$, and $\bm{v}_t^N \coloneqq [v_{tN+N-1}^\top, \dots, v_{tN}^\top]^\top$ denote the stacked process and measurement noise vectors, with means $\mathbf{1}_N \otimes \hat{w}$ and $\mathbf{1}_N \otimes\hat{v}$, and covariances $\Sigma_{w,t}^{N,*}\coloneq \blkdiag(\Sigma_{w,tN+N-1}^*, \dots , \Sigma_{w,tN}^{*})$ and $\Sigma_{v,t}^{N,*}\coloneq \blkdiag(\Sigma_{v,tN+N-1}^*, \dots , \Sigma_{v,tN}^{*})$, respectively.
The downsampled dynamics then take the form
\[
\begin{split}
    x_{t+1}^d &= A^N x_t^d + \mathcal{R}_N \bm{w}_t^N\\
    \bm{y}_t^N &= \mathcal{O}_N x_t^d + \mathcal{H}_N \bm{w}_t^N + \bm{v}_t^N,
\end{split}
\]
where $\bm{y}_t^N \coloneqq [y_{tN+N-1}^\top, \dots, y_{tN}^\top]^\top$ is the stacked observation vector, while the block reachability and observability matrices are defined as
\begin{align}
    \mathcal{R}_N  &\coloneqq 
    \left[
        I_{n_x}, \, A, \,\ldots,\, A^{N-1}
    \right]\\
\mathcal{O}_N  &\coloneqq
    \left[
        (CA^{N-1})^\top,\, \ldots, \,(CA)^\top,\, C^\top
    \right]^\top.
\end{align}

The block Toeplitz matrix $\mathcal{H}_N\in\real{N n_y\times N n_x}$
captures the propagation of process noise into future outputs and is defined elementwise by $[\mathcal{H}_N]_{ij} = CA^{j-i-1}, j>i$, and zero otherwise, for $i,j=1,\ldots,N$. Let $\bm{u}_t^N \coloneq \mathcal{H}_N \bm{w}_t^N + \bm{v}_t^N$. Its covariance is given by $\Sigma_{u,t}^N
=
\mathcal{H}_N \Sigma_{w,t}^{N,*} \mathcal{H}_N^\top
+ \Sigma_{v,t}^{N,*}$.
Adopting the procedure in~\cite{levy2016contraction} and denoting the downsampled prior covariance matrix by $\Xpriord := \Sigma_{x,tN}^-$, the DR Riccati recursion associated with the downsampled model becomes
\begin{equation}\label{eqn:downsampled_Riccati}
\Xpriornextd = r_{\ambset,t}^d(\Xpriord),
\end{equation}
where
\begin{equation}\label{eq:app_block_riccati}
r_{\ambset,t}^d(\Xpriord)
=
\alpha_{N,t}\Bigl((\Xpriord)^{-1} + \Omega_{N,t}\Bigr)^{-1}(\alpha_{N,t})^\top
+
W_{N,t},
\end{equation}
with
\begin{align*}
\alpha_{N,t}&\coloneq A^N - \mathcal{R}_N \mathcal{G}_{N,t} \mathcal{O}_N\\
\Omega_{N,t}
&\coloneq
\mathcal{O}_N^\top (\Sigma_{u,t}^N)^{-1} \mathcal{O}_N\\
W_{N,t}
&\coloneq
\mathcal{R}_N \mathcal{Q}_{N,t} \mathcal{R}_N^\top\\
\mathcal{Q}_{N,t} &\coloneq \Sigma_{w,t}^{N,*} -\Sigma_{w,t}^{N,*} \mathcal{H}_N^\top (\Sigma_{u,t}^N)^{-1} \mathcal{H}_N \Sigma_{w,t}^{N,*}\\
\mathcal{G}_{N,t} &\coloneq \Sigma_{w,t}^{N,*}\mathcal{H}_N^\top (\Sigma_{u,t}^N)^{-1}.
\end{align*}

Under~\Cref{assump:time_inv_nom}, the one-step DR Riccati update is
time-invariant, and the corresponding downsampled mapping
$r_{\ambset,t}^d$ coincides with the $N$-fold iterate $r_{\ambset}^N$ of a
single Riccati operator $r_{\ambset}$.
We therefore analyze contraction properties of the DR Riccati recursion through
this lifted representation, via a downsampling contraction argument in the spirit
of~\cite{zorzi2017convergence}.

We first establish basic structural properties of $\Omega_{N,t}$ and $W_{N,t}$,
which are required for the contraction analysis.

\begin{lemma}
\label{lem:WO_pd}
Suppose~\Cref{assump:obsv,assump:Gauss,assump:time_inv_nom} hold. Then, $W_{N,t}\in\pd{n_x}$ for any $N\geq 1$ and $\Omega_{N,t}\in \pd{n_x}$ for any $N \geq n_x$.
\end{lemma}

The proof is provided in Appendix~\ref{app:WO_pd}.

Using the above properties, we establish contraction of the (downsampled) DR Riccati mapping with respect to Thompson's part metric, defined as follows: for $P,Q\in\pd{n_x}$,
\[
d_T(P,Q) \coloneq \log \left( \max\{\lammax{P^{-1} Q },\lammax{Q^{-1} P}\}\right).
\]
This metric is invariant under congruence transformations and is well suited
for analyzing Riccati-type mappings.

\begin{lemma}\label{lem:contr_factor}
Suppose~\Cref{assump:obsv,assump:Gauss,assump:time_inv_nom} hold and let $N \geq n_x$, so that $\mathcal{O}_N$ has full column rank. Then, for any $t \geq 0$, the downsampled DR Riccati mapping $r_{\ambset,t}^d$~\eqref{eq:app_block_riccati} is strictly contractive on $\pd{n_x}$ with respect to $d_T$: there exists $\xi_{N,t}\in(0,1)$ such that
\[
d_T(r_{\ambset,t}^d(P),r_{\ambset,t}^d(Q))
\le \xi_{N,t}\, d_T(P,Q),\quad \forall P,Q\in\pd{n_x}.
\]
Moreover, the contraction factor satisfies the bound
\begin{equation}\label{eqn:contr_factor}
\xi_{N,t}\le
\left(
\frac{\sqrt{\|M_{N,t}\|_2}}{1+\sqrt{1+\|M_{N,t}\|_2}}
\right)^2,
\end{equation}
where
$M_{N,t}\coloneq
\Omega_{N,t}^{-1/2}\alpha_{N,t}^\top W_{N,t}^{-1}\alpha_{N,t}\Omega_{N,t}^{-1/2}$.
\end{lemma}

This follows from~\cite[Thm.~5.3]{lee2008invariant} together with
\Cref{lem:WO_pd}.

We next show that the contraction factor can be bounded \emph{uniformly} over
time by exploiting the spectral envelope induced by Wasserstein ambiguity.

\begin{proposition}\label{prop:xi_bound}
Suppose that~\Cref{assump:Gauss,assump:time_inv_nom,assump:obsv} hold. Fix
$N\ge n_x$ and define
\begin{equation}\label{eqn:kappa_theta}
\begin{split}
\kappa_N (\theta_v,\theta_w)
\coloneq
\; &\frac{\overline{\lambda}(\hat{\Sigma}_v, \theta_v)
      +\overline{\lambda}(\hat{\Sigma}_w, \theta_w)\|\mathcal{H}_N\|_2^2}
     {\sigma_{\min}(\mathcal{O}_N)^2}\\
&\times
\left(\|A^N\|_2
+ \frac{\overline{\lambda}(\hat{\Sigma}_w, \theta_w)\|\mathcal{H}_N\|_2
        \|\mathcal{R}_N\|_2 \|\mathcal{O}_N\|_2}
       {\lammin{\hat{\Sigma}_v}} \right)^2 
\left(\frac{1}{\lammin{\hat{\Sigma}_w}}
+ \frac{\|\mathcal{H}_N\|_2^2}{\lammin{\hat{\Sigma}_v}}\right),
\end{split}
\end{equation}
where $\overline{\lambda}(\cdot,\cdot)$ is defined as in~\Cref{prop:bures_bound}.
Then, for all $t\ge 0$, the contraction factor in \eqref{eqn:contr_factor}
satisfies
\[
\xi_{N,t}\le \bar\xi_N(\theta_v,\theta_w)
\coloneq
\left(
\frac{\sqrt{\kappa_N(\theta_v,\theta_w)}}
     {1+\sqrt{1+\kappa_N(\theta_v,\theta_w)}}
\right)^2
\in(0,1).
\]
In particular, for any prescribed $\rho\in (0,1)$, any radii $(\theta_v,\theta_w)$ satisfying
\begin{equation}\label{eqn:rho_to_kappa}
\kappa_N(\theta_v,\theta_w)\le \frac{4\rho}{(1-\rho)^2}
\end{equation}
ensure $\xi_{N,t}\le \rho$ for all $t$.
\end{proposition}
The proof is provided in Appendix~\ref{app:xi_bound}.

Combining the pointwise contraction in \Cref{lem:contr_factor} with the uniform
bound in \Cref{prop:xi_bound} yields convergence of the DR Riccati recursion.

\begin{theorem}\label{thm:contraction}
Suppose~\Cref{assump:obsv,assump:Gauss,assump:time_inv_nom} hold and let
$N\ge n_x$. Then, for any $t\ge 0$, the downsampled DR Riccati mapping
$r_{\ambset,t}^d$ is strictly contractive on $\pd{n_x}$ with respect to
Thompson's part metric, with contraction factor $\xi_{N,t}$ satisfying
\[
\xi_{N,t}\le \bar\xi_N(\theta_v,\theta_w)<1 \qquad \forall t\ge 0.
\]
Consequently, the full DR Riccati recursion~\eqref{eqn:riccati_like}
converges to a unique fixed point $\Xpriorss\succ 0$, and the associated DR
Kalman gain $K_t$ converges.
\end{theorem}
The proof is provided in Appendix~\ref{app:contraction}.

As a consequence of~\Cref{thm:contraction}, the steady-state solution can be
obtained by solving the single stationary SDP~\eqref{eqn:DRKF_SDP_ss} offline,
rather than via online Riccati iterations.

\begin{corollary}\label{cor:ss_sdp}
Under the assumptions of~\Cref{thm:contraction}, the steady-state covariance
matrices $(\Xpriorssopt,\Xpostssopt)$, together with the least-favorable noise
covariances $\Sigma_{w,\infty}^*$ and $\Sigma_{v,\infty}^*$, solve the
stationary SDP~\eqref{eqn:DRKF_SDP_ss}. Consequently,
$K_\infty^* = \lim_{t\to\infty} K_t$.
\end{corollary}
The proof can be found in Appendix~\ref{app:ss_sdp}.

Under~\Cref{assump:obsv}, the Riccati recursions of the standard KFs in
\Cref{thm:drkf_sandwich} converge. 
Taking limits in the matrix inequalities $A\preceq B$ (in the sense of the
Loewner partial order) appearing in \Cref{thm:drkf_sandwich}, and using that the
positive semidefinite cone is closed, yields the following steady-state bounds.

\begin{corollary}\label{cor:ss_sandwich}
Under the assumptions of~\Cref{thm:contraction}, the steady-state prior and
posterior error covariances $\Xpriorss$ and $\Xpostss$ satisfy
\[
\Xpriorlowss \preceq \Xpriorss \preceq \Xpriorhighss,
\qquad
\Xpostlowss \preceq \Xpostss \preceq \Xposthighss,
\]
where $\Xpriorlowss\coloneq \lim_{t\to\infty}\Xpriorlow$ and
$\Xpriorhighss\coloneq \lim_{t\to\infty}\Xpriorhigh$ (and similarly for
$\Xpostlowss,\Xposthighss$) are the unique steady-state limits of the lower- and
upper-bound KF recursions in~\Cref{thm:drkf_sandwich}.
\end{corollary}

\begin{remark}
Since $\overline{\lambda}(\hat X,\theta)$ is nondecreasing in $\theta$  for $\theta \ge 0$, the bound
$\kappa_N(\theta_v,\theta_w)$ in~\eqref{eqn:kappa_theta} is monotone in both
ambiguity radii. Consequently, a practical tuning procedure for
$(\theta_v,\theta_w)$ consists of enforcing the rate constraint
\eqref{eqn:rho_to_kappa} using one of the following strategies:
(i) fix $\theta_v$ and determine the largest admissible $\theta_w$;
(ii) fix $\theta_w$ and determine the largest admissible $\theta_v$; or
(iii) restrict $(\theta_v,\theta_w)$ to a one-dimensional curve
(e.g., $\theta_w=\gamma\theta_v$ for some $\gamma>0$) and select the maximal
radii satisfying~\eqref{eqn:rho_to_kappa}.
These procedures allow one to balance robustness and convergence speed in a
transparent and computationally tractable manner.
\end{remark}

\subsection{Stability}

A key property of the classical Kalman filter is the stability of its
steady-state estimation error dynamics: under detectability of $(A,C)$ and
positive definite process noise, the closed-loop error matrix $(I-KC)A$ is
Schur stable, ensuring bounded error covariance and asymptotic convergence of
the estimation error in the unbiased case.

In the distributionally robust  setting, however, the estimator is designed
against least-favorable noise distributions within ambiguity sets. It is thus
nontrivial whether the resulting DRKF, whose steady-state
gain is characterized by the minimax design in \Cref{thm:contraction}, preserves
the stability guarantees of the classical KF. The following result shows that,
despite the adversarial nature of the noise model, the steady-state DRKF remains
asymptotically stable in the same sense as the classical KF.

\begin{theorem}\label{thm:drkf_stability}
Suppose~\Cref{assump:obsv,assump:Gauss,assump:time_inv_nom} hold. Let $(\Xpriorssopt,\Xpostssopt)$ and $(\Sigma_{w,\infty}^*,\Sigma_{v,\infty}^*)$ denote the steady-state DRKF covariances, and let $K_\infty^*$ be the corresponding steady-state DR Kalman gain. Define the closed-loop error matrix $F_\infty \coloneq (I - K_\infty^* C) A$.
Then, $F_\infty$ is Schur stable, and the posterior estimation error
$e_t := x_t - \xpost_t$ evolves according to
\[
e_t \;=\; F_\infty e_{t-1} + (I-K_\infty^* C)\,(w_{t-1}-\hat{w}) - K_\infty^* (v_t-\hat{v}).
\]
Consequently, under the least-favorable noise distributions, the estimation error is mean-square stable with
\[
\sup_{t\ge 0}\,\mathbb{E}[\|e_t\|^2] \;<\;\infty, \quad \text{and} \quad \lim_{t\to\infty} \mathrm{cov}(e_t) = \Xpostssopt.
\]
\end{theorem}

The proof is provided in Appendix~\ref{app:drkf_stability}.

While~\Cref{thm:drkf_stability} establishes mean-square stability, it does not by
itself characterize the behavior of the estimation-error mean under model
mismatch. In practice, the true noise processes may have means that differ from the
nominal values used by the filter, leading to a steady-state bias. The following corollary characterizes this bias under the true noise
distributions.

\begin{corollary}\label{cor:drkf_error_bias}
Under the assumptions of~\Cref{thm:drkf_stability}, consider running the
steady-state DRKF with gain $K_\infty^*$ on a system whose noise processes have
constant (true) means $\mathbb{E}[w_t]=\mu_w$ and $\mathbb{E}[v_t]=\mu_v$ for all
$t$, with finite first moments.
Let $m_t := \mathbb{E}[e_t]$ denote the posterior error mean under the true noise
distributions.
Then,
$m_t = F_\infty\, m_{t-1} + (I-K_\infty^* C)(\mu_w - \hat{w}) - K_\infty^* (\mu_v - \hat{v})$,
and hence
\[
\lim_{t\to\infty} m_t
= (I - F_\infty)^{-1} \bigl((I-K_\infty^* C)(\mu_w - \hat{w}) - K_\infty^* (\mu_v - \hat{v})\bigr).
\]
In particular, if the nominal noise means used by the filter coincide with the true noise means ($\hat{w}=\mu_w$ and $\hat{v}=\mu_v$), then $\mathbb{E}[e_t]\to 0$.
\end{corollary}

The proof is provided in Appendix~\ref{app:drkf_error_bias}.

As shown in \Cref{lem:GelbrichMMSE}, mean shifts are suboptimal for the adversary
under Wasserstein ambiguity at the design stage, so the least-favorable noise
distributions preserve the nominal means. Therefore, when the nominal noise means used by the filter
coincide with the true noise means, the steady-state DRKF is unbiased in
operation. When this condition is violated, a constant steady-state bias
appears, exactly as characterized in \Cref{cor:drkf_error_bias}.

\subsection{Optimality}

Having established convergence and stability of the proposed DRKF, we now
address its performance optimality. In the presence of distributional
uncertainty, a natural question is whether one can design a causal estimator
that outperforms the DRKF in terms of worst-case estimation accuracy.

In this subsection, we show that this is not possible asymptotically: among all
causal estimators, the steady-state DRKF minimizes the worst-case one-step
conditional mean-square error in the limit, as well as its long-run time
average.

\begin{theorem}\label{thm:optim}
Suppose~\Cref{assump:Gauss,assump:time_inv_nom,assump:obsv} hold.
Let $\Xpriorssopt \in \pd{n_x}$ denote the unique fixed point of the DR Riccati map $r_{\ambset}$ and let $\Xpostssopt$ be the corresponding steady-state posterior \emph{error} covariance.
Then, for any causal state estimator sequence $\{\psi_t\}_{t\ge 0}$ with $\psi_t \in \mathcal{F}_t$, the following inequalities hold almost surely with
respect to $\sigma(\mathcal{Y}_{t-1})$:
\begin{align}
    \liminf_{t \to \infty}  \sup_{\Pdist_{e,t} \in \ambset_{e,t}} J_t(\psi_t, \Pdist_{e,t})
    &\geq  \Tr[\Xpostssopt],\label{eqn:lower_bound_liminf}\\
    \liminf_{T \to \infty} \ \frac{1}{T} \sum_{t=0}^{T-1}  
    \sup_{\Pdist_{e,t} \in \ambset_{e,t}}
    J_t(\psi_t, \Pdist_{e,t})
    &\geq \Tr[\Xpostssopt].\label{eqn:lower_bound_avg}
\end{align}
Moreover, the steady-state DRKF $\psi_\infty$ with gain $K_\infty^*$ achieves these bounds with equality:
\begin{align}
    \lim_{t \to \infty} \ \sup_{\Pdist_{e,t} \in \ambset_{e,t}}
    J_t(\psi_\infty, \Pdist_{e,t})
    &= \Tr[\Xpostssopt], \label{eq:psi_infty_stage}\\
    \lim_{T \to \infty} \ \frac{1}{T} \sum_{t=0}^{T-1} \ 
    \sup_{\Pdist_{e,t} \in \ambset_{e,t}}
    J_t(\psi_\infty, \Pdist_{e,t})
    &= \Tr[\Xpostssopt]. \label{eq:psi_infty_cesaro}
\end{align}
Hence, the steady-state DRKF is asymptotically minimax-optimal with respect to the worst-case one-step and long-run average MSE.
\end{theorem}

The proof of is provided in Appendix~\ref{app:optim}. This result shows that the steady-state DRKF attains the smallest possible worst-case asymptotic MSE among all causal estimators, thereby extending the fundamental optimality of the classical KF to the DR setting. Importantly, the theorem concerns
\emph{asymptotic} performance: establishing finite-horizon minimax optimality
under stage-wise ambiguity leads to a substantially more intricate dynamic game
and is beyond the scope of this work.

\begin{remark}
When $\theta_v=\theta_w=0$, the ambiguity sets reduce to the nominal noise
distributions. In this case, the least-favorable covariances coincide with the
nominal ones, and the DRKF reduces exactly to the classical KF. Thus,
\Cref{thm:optim} is consistent with the classical KF optimality and recovers it
as a special case in the asymptotic regime.
\end{remark}

\section{Experiments}\label{sec:simulations}
\begin{figure*}[ht!]
    \centering
    \includegraphics[width=0.95\linewidth]{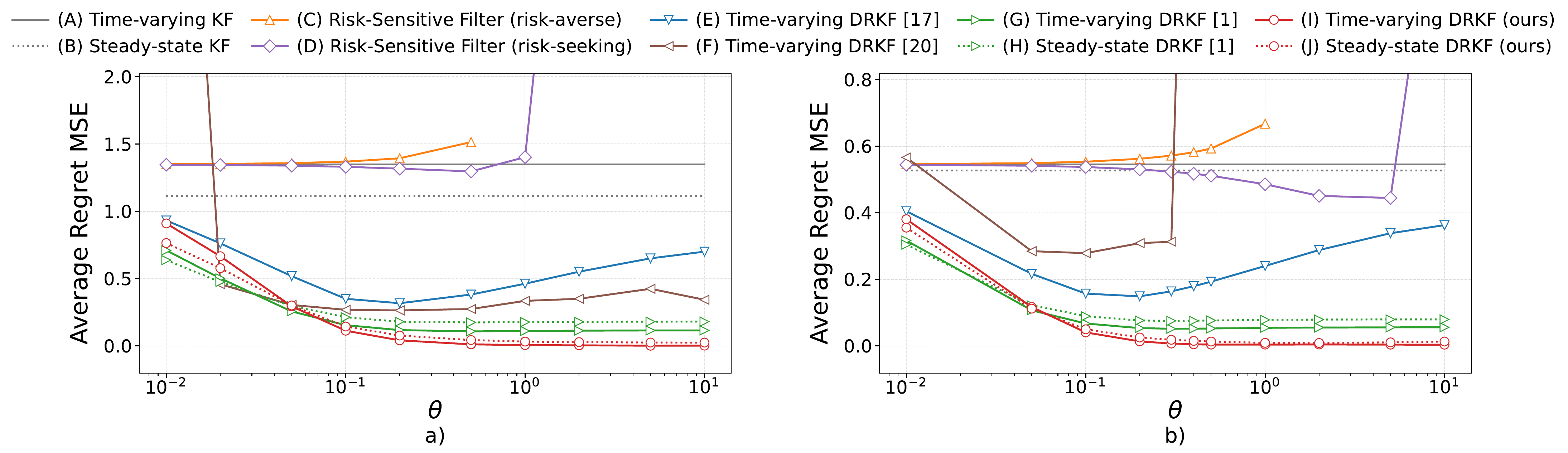}
    \caption{Effect of $\theta$ on the average regret MSE under a) Gaussian and b) U-Quadratic noise distributions, averaged over 20 simulation runs.}
    \label{fig:theta_effect_regret}
\end{figure*}

We evaluate the proposed DRKFs through numerical experiments that assess estimation accuracy,
closed-loop performance, and computational efficiency, and that empirically validate the
theoretical results in~\Cref{sec:theoretical}.\footnote{All experiments were performed on a laptop equipped with an Intel Core Ultra 7 155H @ 3.80\,GHz and 32\,GB of RAM. Source code is available at \url{https://github.com/jangminhyuk/DRKF2025}.}

\subsection{Comparison of Robust State Estimators}\label{subsec:filter_compare}

We begin by comparing the proposed time-varying and steady-state DRKFs with the following baseline
estimators:
\begin{methodlist}
  \item \label{m:TVKF} Standard time-varying KF
  \item \label{m:SSKF} Steady-state KF
  \item \label{m:RSF} Risk-sensitive filter (risk-averse)~\cite{whittle1981risk,jacobson1973optimal,levy2016contraction,zorzi2017convergence} 
  \item \label{m:RSF2} Risk-sensitive filter (risk-seeking)~\cite{whittle1981risk,jacobson1973optimal,levy2016contraction,zorzi2017convergence} 
  \item \label{m:DRKF-NeurIPS} DRKF with joint Wasserstein ambiguity on  $(x_t, y_t)$~\cite{NEURIPS_DRKF}
  \item \label{m:BCOT} DRKF with bicausal Wasserstein ambiguity~\cite{han2024distributionally}\footnote{Following~\cite{han2024distributionally}, we cap the number of internal iterations at 20.}
  \item \label{m:DRKF-JangThm1} Time-varying DRKF with Wasserstein ambiguity on the prior state and measurement noise~\cite[Thm.~1]{jang2025steady}
  \item \label{m:SSDRKF-JangSec4} Time-invariant version of~\mref{m:DRKF-JangThm1} with stationary nominals~\cite[Sect.~4]{jang2025steady}
  \item \label{m:DRKF-Ours} \textbf{Time-varying DRKF (ours)} [\Cref{alg:DRKF}] 
  
  \item \label{m:SSDRKF-Ours} \textbf{Steady-state DRKF (ours)} [\Cref{alg:SSDRKF}]
   
\end{methodlist}

We use the same symbol $\theta$ to denote the robustness parameter across all methods, although
its interpretation differs.
For risk-sensitive filters~\mref{m:RSF} and~\mref{m:RSF2}, $\theta$ is the risk parameter (with opposite signs for
risk-averse/seeking variants), chosen within the stability range~\cite{levy2016contraction}.
For Wasserstein-based methods, $\theta$ denotes the ambiguity radius.
 For \mref{m:DRKF-JangThm1} and \mref{m:SSDRKF-JangSec4}, we set $\theta_{x_0} = \theta_v = \theta$. In our proposed formulations, we set $\theta_{x_0}=\theta_w=\theta_v=\theta$ for \mref{m:DRKF-Ours}, and $\theta_w=\theta_v=\theta$ for \mref{m:SSDRKF-Ours}. To ensure fairness, we sweep each method over its admissible range of $\theta$ and report
performance at the best tuned value.\footnote{In practice, DRKF radii $\theta$ can be selected from data, following standard approaches in the DRO literature, such as cross-validation or bootstrapping (e.g.,~\cite{esfahani2015data,gao2024wasserstein}).}

\subsubsection{Estimation Accuracy}\label{subsub:estimation}

We first consider a linear time-invariant system, subject to inaccuracies in both the process and measurement noise distributions:
\[
\begin{split}
A
&=  \begin{bmatrix}
    1 & 0.2  & 0 & 0 \\
    0 & 0.2  & 0.2 & 0\\
    0 & 0 & 0.2 & 0.2\\
    0 & 0 & 0 & -1
\end{bmatrix}, \quad C = 
\begin{bmatrix} 1 & 0 & 0 &0\\
0 & 0 & 1 & 0\end{bmatrix}.
\end{split}
\]
We test two noise models: $(i)$ Gaussian, with $x_0 \sim \mathcal{N}(0, 0.01 I_4)$, $w_t \sim \mathcal{N}(0, 0.01 I_4)$, $v_t \sim \mathcal{N}(0, 0.01 I_2)$, and $(ii)$ U-Quadratic, with $x_0 \sim \mathcal{UQ}([-0.1, 0.1]^4)$, $w_t \sim \mathcal{UQ}([-0.1, 0.1]^4)$, $v_t \sim \mathcal{UQ}([-0.1, 0.1]^2)$.
Nominal covariances are learned from only 10 measurement samples using the expectation-maximization (EM) method in~\cite{han2024distributionally}, creating a significant mismatch between nominal and true noise statistics.
The estimation horizon is $T=50$.

\Cref{fig:theta_effect_regret} shows the effect of $\theta\in[10^{-2},10^{1}]$ on the average
regret MSE under both noise models, averaged over $20$ runs.
The regret MSE is defined as the difference between the filter's MSE and that of the KF using the \emph{true} noise distributions. 
The results show that both proposed DRKFs consistently achieve the lowest tuned regret and remain competitive across a wide range of $\theta$. In particular, \mref{m:DRKF-Ours} achieves performance closest to the optimal estimator under both poorly estimated Gaussian and non-Gaussian noise.

In contrast, the BCOT-based DRKF~\mref{m:BCOT} exhibits numerical instability and discontinuous
performance across $\theta$, even when increasing the internal iteration limit to 50.
 The risk-sensitive filters~\mref{m:RSF} and~\mref{m:RSF2} become overly conservative for large $\theta$, leading to degraded accuracy.

\begin{figure}[t]
    \centering
\includegraphics[width=0.7\linewidth]{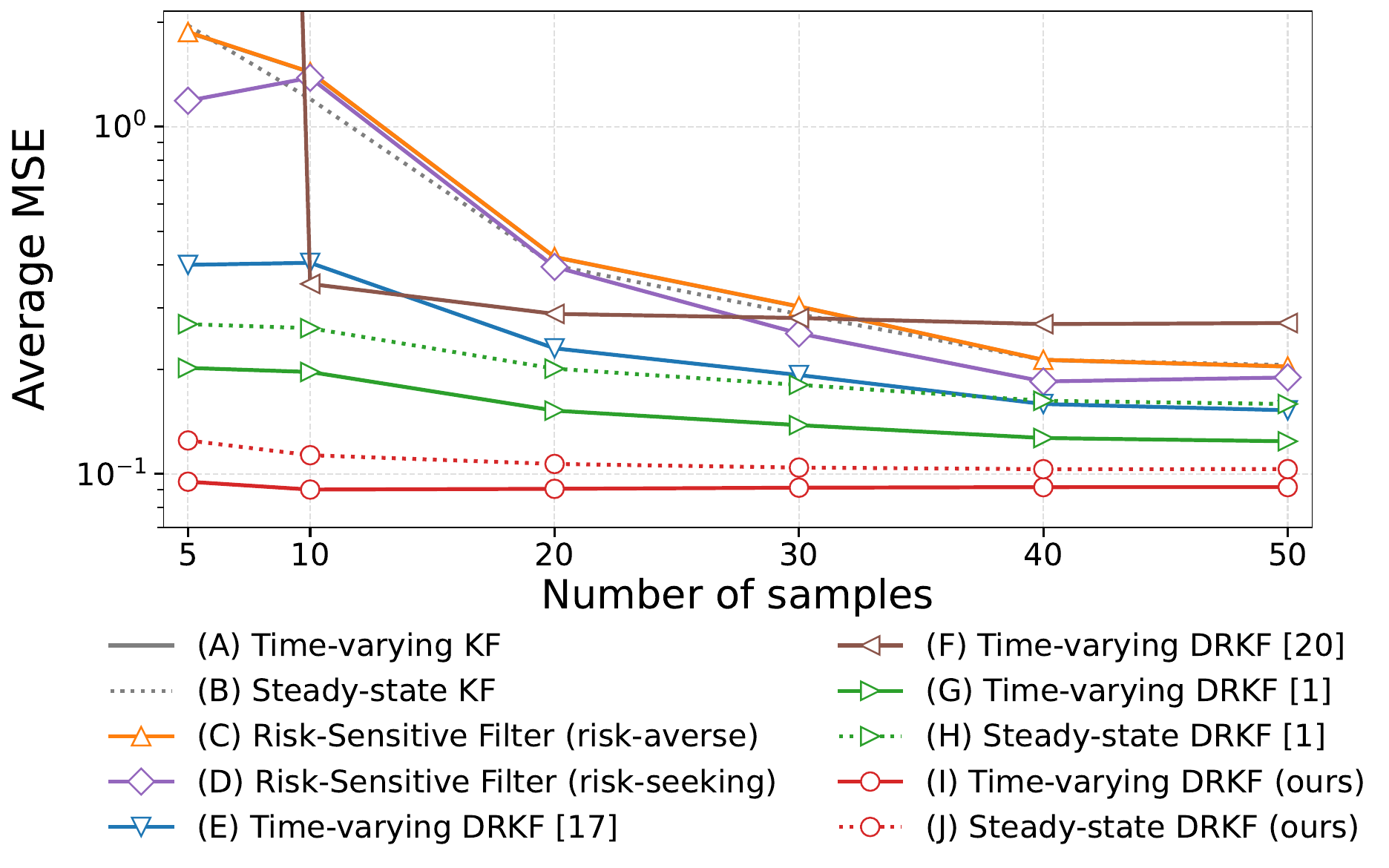}
    \caption{Effect of the number of samples used to construct the nominal distributions on the average MSE under Gaussian noise, averaged over 10 runs.}
    \label{fig:datasize}
\end{figure}

\begin{figure}[t]
    \centering \includegraphics[width=0.8\linewidth, trim=0cm 3cm 0cm 4.5cm, clip]{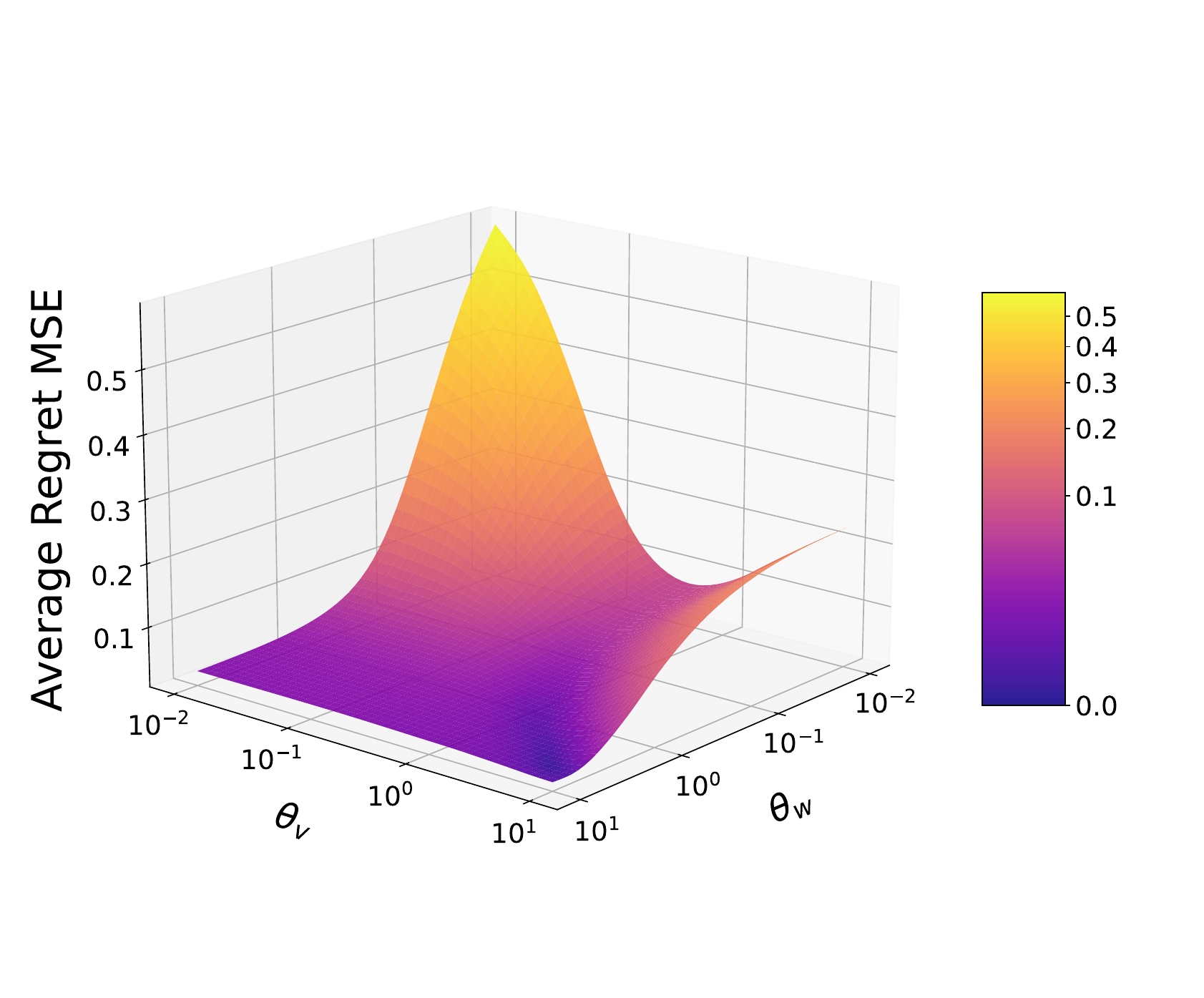}
    \caption{Average regret MSE of the proposed steady-state DRKF under Gaussian noise, as a function of $\theta_w$ and $\theta_v$, averaged over 100 runs.}
\label{fig:theta_wv_3d}
\end{figure}

\begin{figure}[t!]
    \centering
\includegraphics[width=\linewidth]{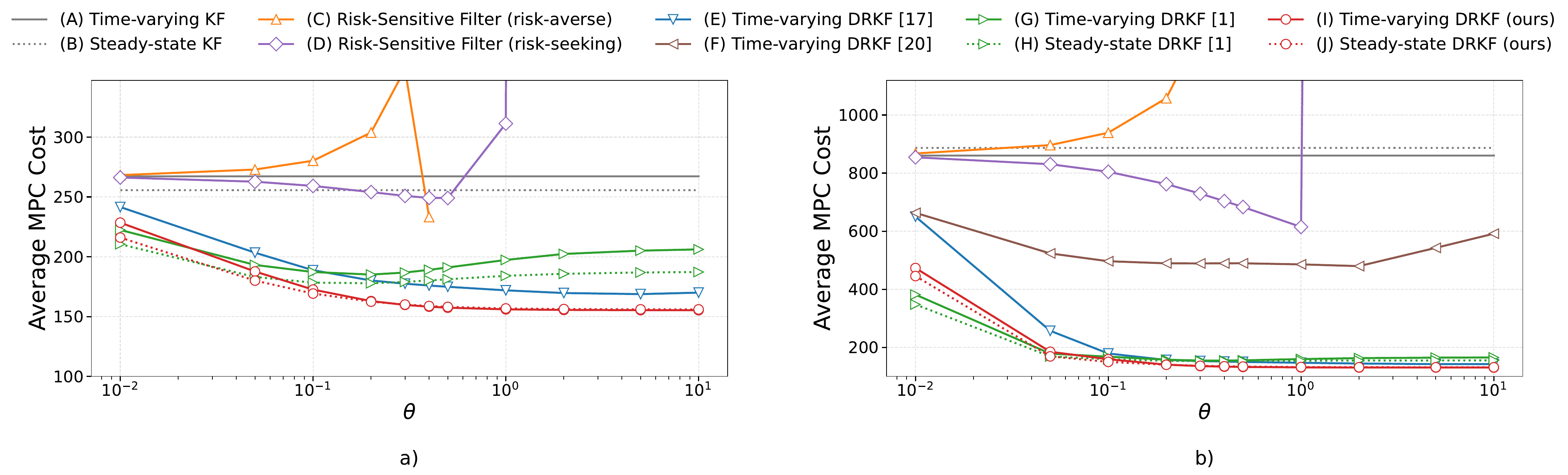}
    \caption{Effect of $\theta$ on the average MPC cost under a) Gaussian and b) nonzero-mean U-Quadratic noise, averaged over 20 simulation runs.} 
    \label{fig:traj_mpc}
\end{figure}

\begin{figure*}[t!]
    \centering
\includegraphics[width=\linewidth]{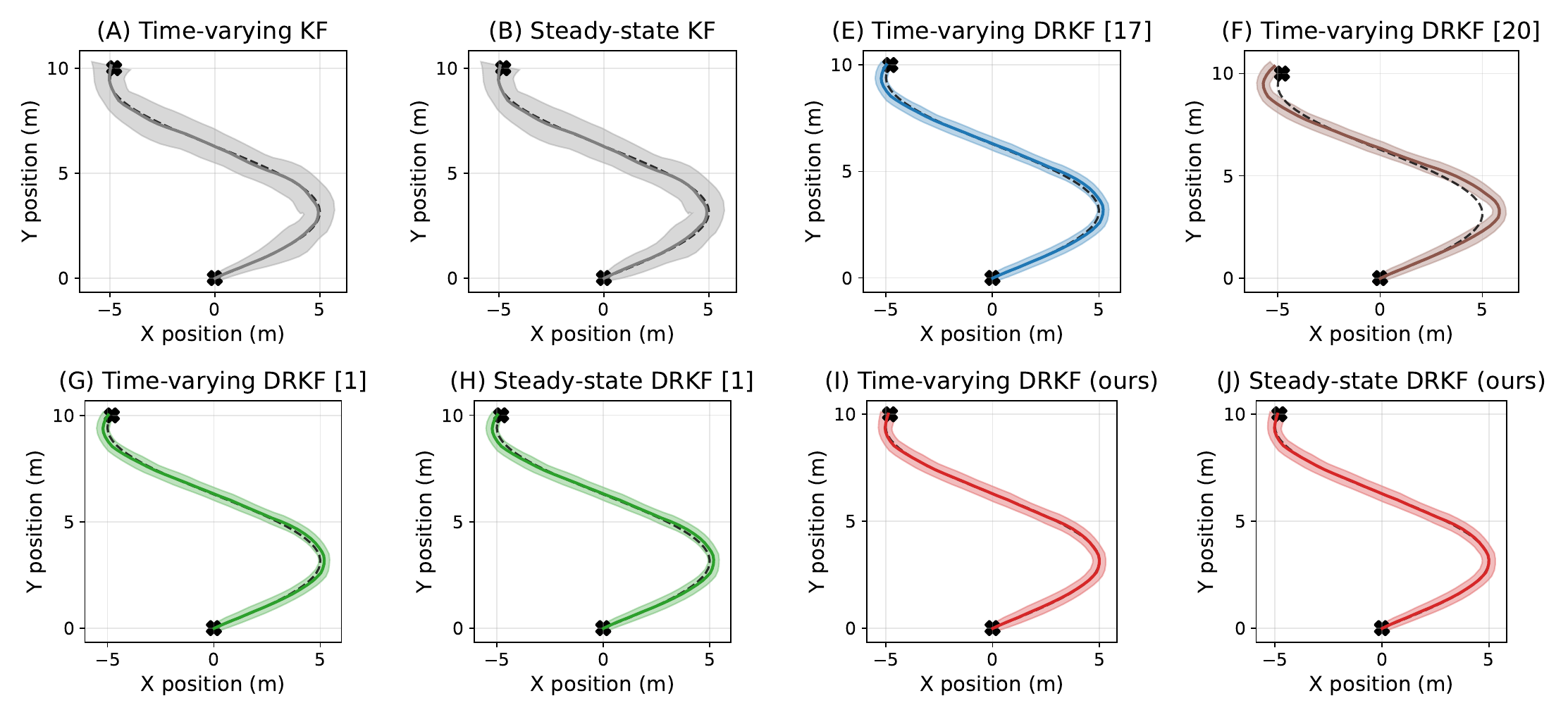}
    \caption{2D trajectories averaged over 200 simulations. Each filter uses its optimal $\theta \in [10^{-2}, 10^{1}]$ minimizing the MPC cost. The dashed black line is the desired trajectory; colored curves show the mean, and shaded tubes indicate $\pm 1$ standard deviation.}
    \label{fig:traj_subplot}
\end{figure*}

\Cref{fig:datasize} further examines performance as a function of the number of samples used to
estimate the nominal noise statistics.\footnote{Each filter is optimally tuned over $\theta \in [10^{-2},10^{1}]$, and we vary the number of input-output samples used in the EM procedure so that small sample sizes correspond to less accurate nominal distributions.} 
Across all sample sizes, the proposed DRKFs achieve the lowest MSE, with particularly clear
advantages in the low-data regime.

\Cref{fig:theta_wv_3d} illustrates the sensitivity of the steady-state DRKF to the ambiguity
radii $(\theta_w,\theta_v)$.
Insufficient robustness (small radii) leads to large regret, while overly large radii induce
excessive conservatism, highlighting the need to balance $\theta_w$ and $\theta_v$.

\subsubsection{Trajectory Tracking Task}
\begin{table}[t!]
    \centering
        \caption{Mean and standard deviation of the acceleration-command magnitude $\|u_t\|_2$ for each filter, averaged over 200 simulations.}
    \label{tab:fullwidth-9col}
    \setlength{\tabcolsep}{2.5pt}
\begin{tabular}{l*{8}{>{\centering\arraybackslash}m{1.7cm}}}
\toprule
 & \mref{m:TVKF} & \mref{m:SSKF} & \mref{m:DRKF-NeurIPS} & \mref{m:BCOT} & \mref{m:DRKF-JangThm1} & \mref{m:SSDRKF-JangSec4} & \mref{m:DRKF-Ours} & \mref{m:SSDRKF-Ours} \\
\midrule
$\|u_t\|_2$
& 5.69 (7.44) & 5.91 (7.70) & 3.29 (1.93) & 2.70 (1.78) & 3.37 (2.02) & 3.30 (1.99) & \textbf{2.18} \textbf{(1.63)} & \textbf{2.20} \textbf{(1.64)} \\
\bottomrule
\end{tabular}\label{table:input}
\end{table}

We next evaluate closed-loop performance on a 2D trajectory-tracking task, following the setup
in~\cite[Sect.~5.2]{jang2025steady}:
\[
\begin{split}
x_{t+1} 
&=  \begin{bmatrix}
    I_2 & \Delta t I_2 \\ \mathbf{0}_{2\times 2} & I_2
\end{bmatrix} x_t + \begin{bmatrix}
    0.5 (\Delta t)^2 I_2 \\ \Delta t I_2
\end{bmatrix} u_t + w_t\\
y_t &=  
\begin{bmatrix} I_2 &  \mathbf{0}_{2\times 2} \end{bmatrix}  x_t + v_t,
\end{split}
\]
where $\Delta t = 0.2\,\text{s}$. The state $x_t = [p_t^x, p_t^y, v_t^x, v_t^y]^\top$ collects the planar position and velocity, and the control input $u_t = [a_t^x, a_t^y]^\top$ represents acceleration commands.
All methods use the same observer-based MPC controller, with the current state estimate $\xpost_t$ provided by each filter. {The controller solves a finite-horizon tracking problem with horizon $10$ and weights $Q = \mathrm{diag}([10,10,1,1])$ and $R = 0.1 I_2$, where $Q$ penalizes tracking error and $R$ penalizes control effort. The first control input $u_t$ is applied, and the optimization is repeated at the next step using $\xpost_{t+1}$.} 

The true noise distributions are $x_0 \sim \mathcal{N}(0, 0.02 I_4)$, $w_t \sim \mathcal{N}(0, 0.02 I_4)$, and $v_t \sim \mathcal{N}(0, 0.02 I_2)$, simulated over $10\,\text{s}$. For the nominal distributions, we fit an EM model using only $1\,\text{s}$ of input-output data, yielding noticeably inaccurate nominal statistics and reflecting the practical challenge of estimating noise models from limited observations.

\begin{figure}[t!]
    \centering \includegraphics[width=0.6\linewidth]{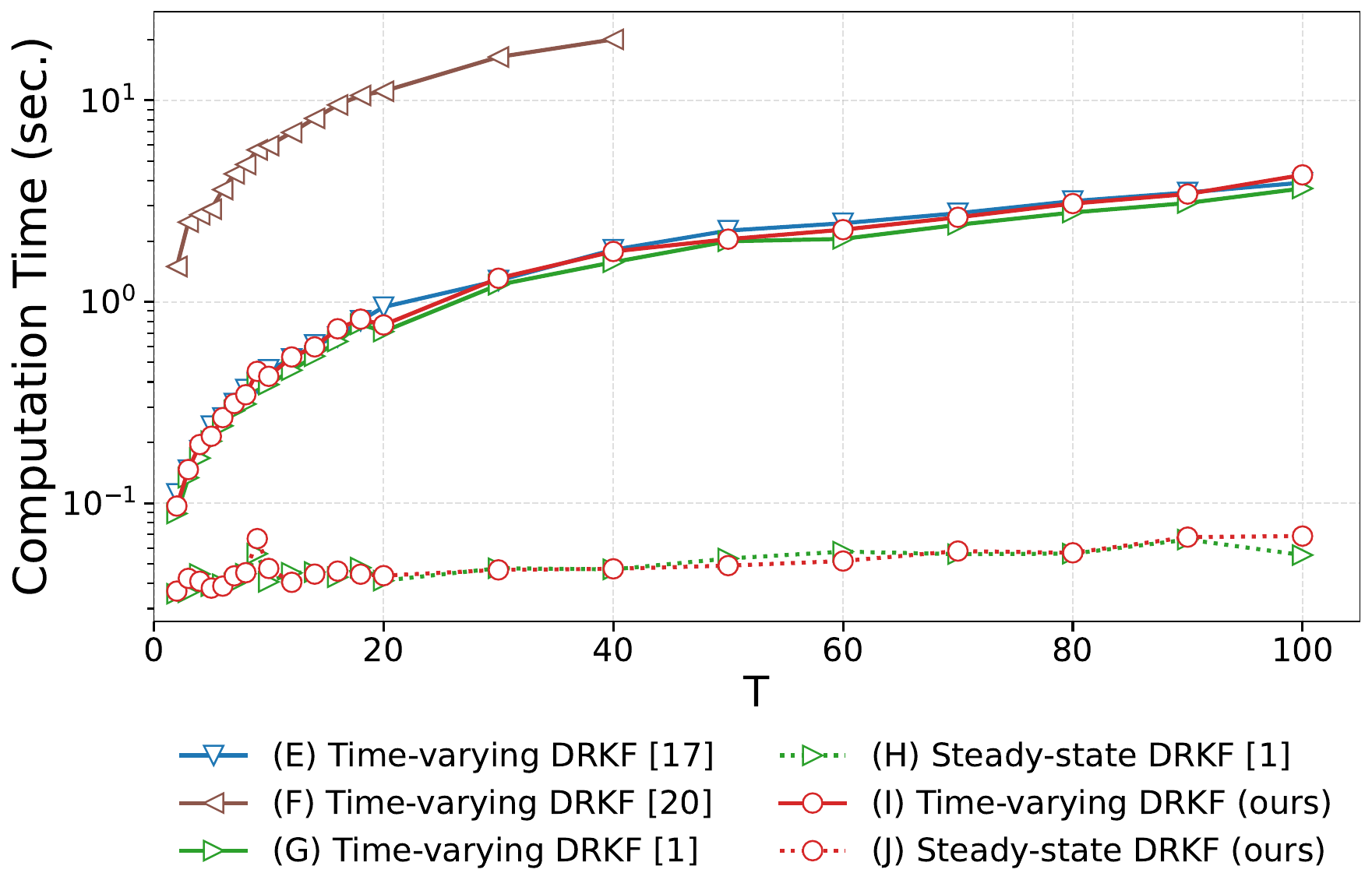}
    \caption{Average offline computation time as a function of the horizon length $T$ for $n_x=4$, $n_y=2$, averaged over 10 runs.}
\label{fig:time_horizon}
\vspace{-0.1in}
\end{figure}
\begin{table}[t]
\centering
\caption{Average offline computation time of the steady-state DRKF as a function of $n_x = n_y = n$ (10 runs).}
\label{tab:ss_drkf_time}
\renewcommand{\arraystretch}{1.2}
\setlength{\tabcolsep}{6.0pt}  
\begin{tabular}{c c c c c c c c c}
\hline
$\bm{n}$ & \textbf{5} & \textbf{10} & \textbf{15} & \textbf{20} & \textbf{25} & \textbf{30} & \textbf{35} & \textbf{40} \\
\hline
Time (s) 
& 0.0600 & 0.1794 & 0.4736 & 1.2414
& 2.2905 & 4.8900 & 8.3890 & 12.4232 \\
\hline
\end{tabular}
\end{table}

\Cref{fig:traj_mpc} shows the effect of $\theta$ on the average MPC cost, defined as the closed-loop quadratic sum of tracking error and input energy weighted by $Q$ and $R$. Consistent with~\Cref{subsub:estimation}, both our time-varying and steady-state DRKFs achieve the lowest cost when optimally tuned, even in the presence of non-Gaussian disturbances and nonzero-mean noise.

\Cref{fig:traj_subplot} displays the mean trajectories and uncertainty envelopes obtained with the best-tuned radius $\theta \in [10^{-2},10^{1}]$ for each filter. The standard KFs~\mref{m:TVKF} and~\mref{m:SSKF} produce wider uncertainty tubes, indicating higher estimation variance. 
The BCOT-based DRKF~\mref{m:BCOT} reduces dispersion but introduces greater bias, especially during sharp turns. 
Other DRKFs--\mref{m:DRKF-NeurIPS}, \mref{m:DRKF-JangThm1}, and \mref{m:SSDRKF-JangSec4}--achieve improved tracking, while the proposed time-varying and steady-state DRKFs
exhibit the lowest dispersion and bias.

Although mean trajectories are similar across DRKFs, control effort differs substantially.
As shown in \Cref{table:input}, the proposed DRKFs require the lowest average control energy among
all methods.

\subsubsection{Computation Time}

\Cref{fig:time_horizon} compares the offline computation time of \mref{m:DRKF-NeurIPS}--\mref{m:SSDRKF-Ours} as a function of the horizon length $T$.\footnote{We use the same setup as in~\Cref{subsub:estimation} with $\theta=1.0$ and Gaussian noise.}
Time-varying DRKFs exhibit linear growth in offline computation time with respect to $T$, with \mref{m:BCOT} being particularly expensive. 
In contrast, the steady-state DRKFs \mref{m:SSDRKF-JangSec4} and \mref{m:SSDRKF-Ours} maintain constant runtime, as they require solving only a single
SDP offline.
This demonstrates the scalability of the proposed steady-state DRKF for long-horizon
applications.

Finally,~\Cref{tab:ss_drkf_time} evaluates scalability with respect to the system dimension. For each $n$, we generate a random system with $n_x = n_y = n$ and measure the offline computation time. Even at $n = 40$, the total runtime remains under $13$ seconds, confirming the practical efficiency of the steady-state DRKF for moderately sized systems.

\subsection{Validation of Theoretical Properties}\label{subsec:exp_theoretical}

\begin{figure}[t!]
    \centering
    \includegraphics[width=0.6\linewidth]{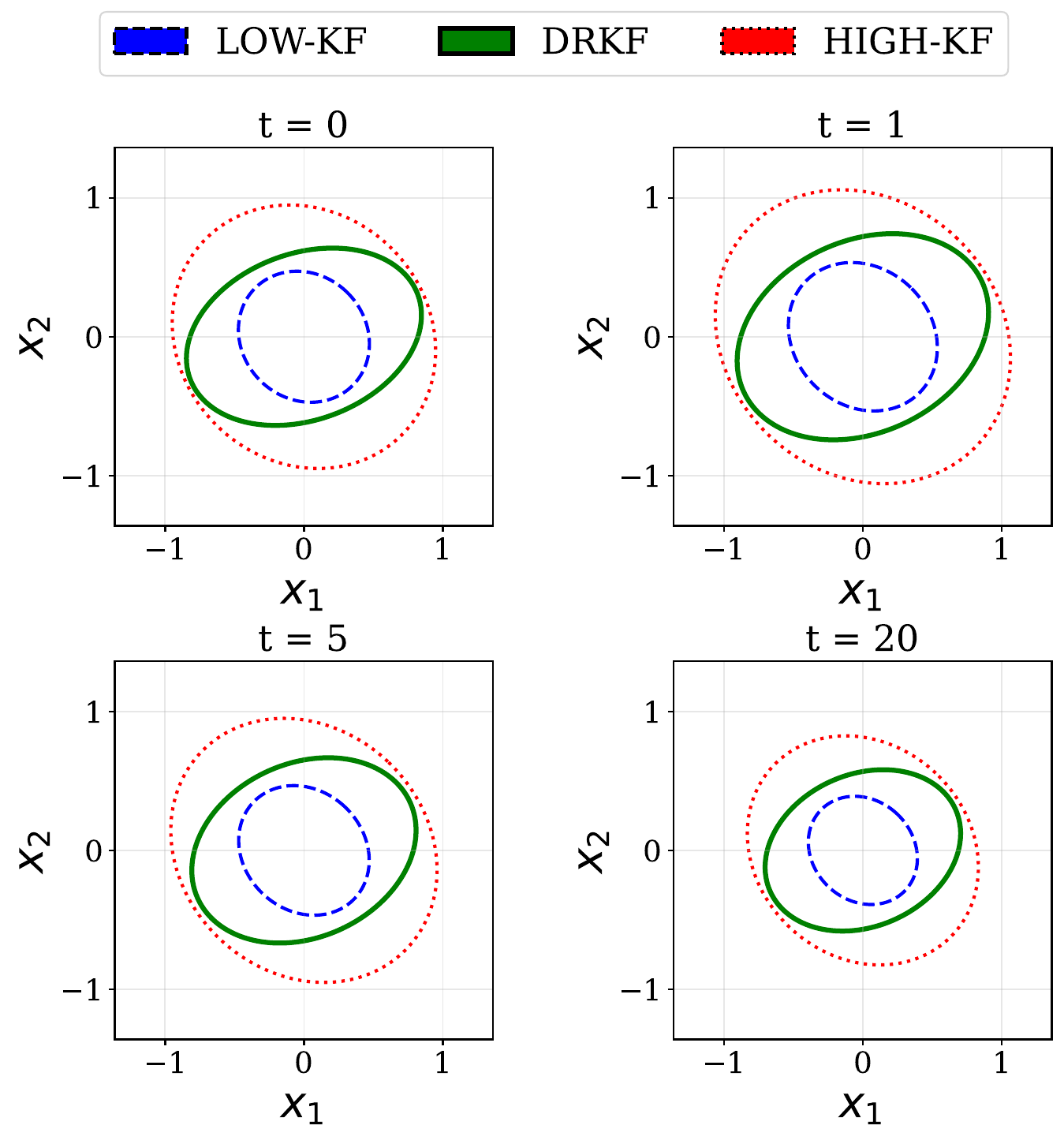}
    \caption{95\% confidence ellipses of the posterior covariances for LOW-KF, DRKF, and HIGH-KF at $t\in\{0,1,5,20\}$.}
    \label{fig:ellipses_2D}
\end{figure}
\begin{figure}[t!]
    \centering
\includegraphics[width=0.6\linewidth,clip,trim=1cm 3.0cm 0cm 3cm]{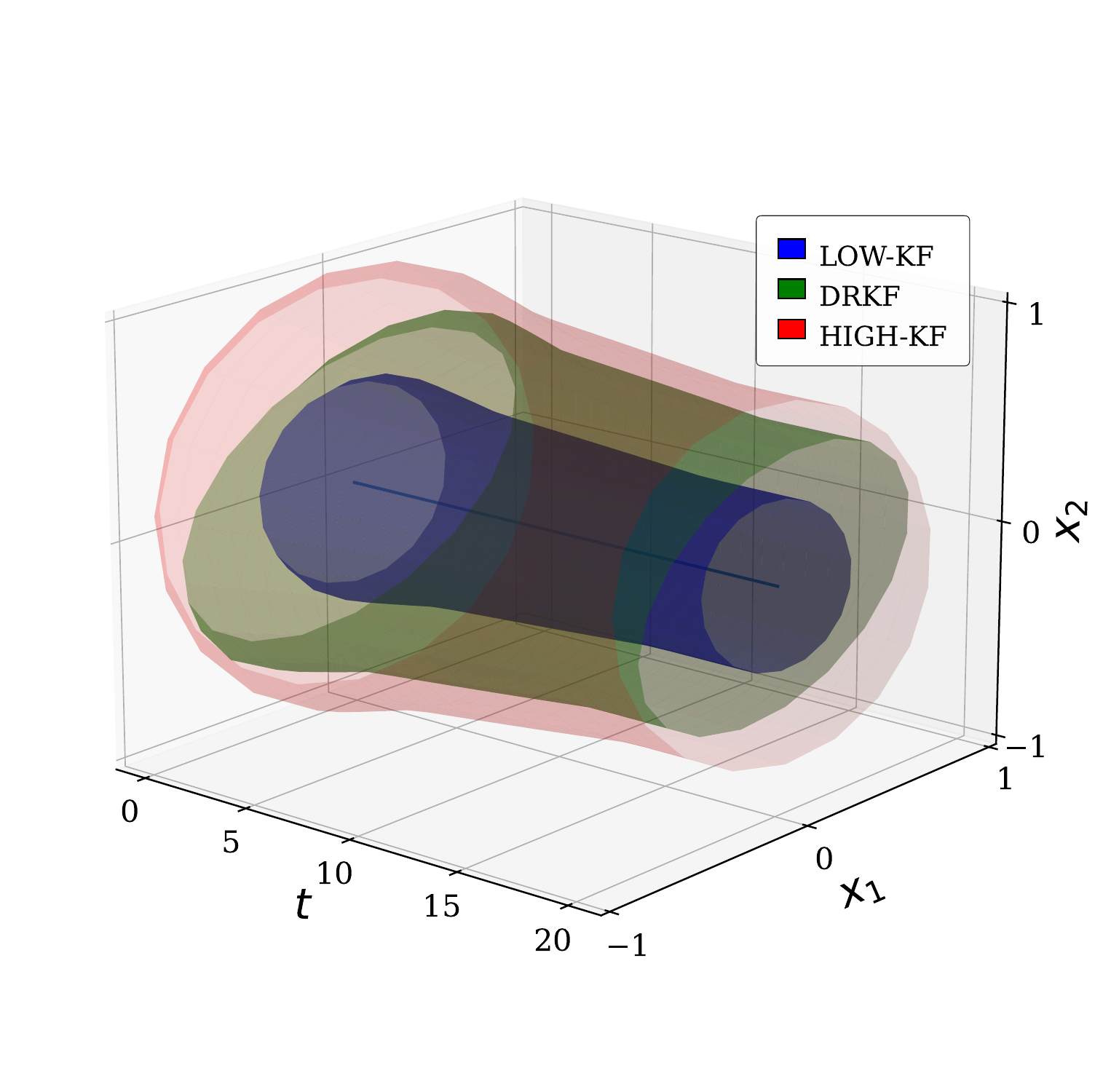}
    \caption{Three-dimensional tube visualization obtained by stacking the $95\%$ ellipses of the posterior covariances of LOW-KF, DRKF, and HIGH-KF for $t=0,\dots,20$.}
    \label{fig:tube_3D}
\end{figure}

\subsubsection{Covariance Sandwich Property}\label{subsubsec:exp_sandwich}

We illustrate the covariance sandwich property of~\Cref{thm:drkf_sandwich} using a 2D example with
\[
A = 0.95\begin{bmatrix}\cos(\pi/8) & -\sin(\pi/8)\\ \sin(\pi/8) & \cos(\pi/8)\end{bmatrix}, \qquad 
C = \begin{bmatrix}1.0 & 0.1\\ 0.1 & 1.0\end{bmatrix},
\]
horizon $T=20$, and ambiguity radii $\theta_w=\theta_v=\theta_{x_0}=0.1$.
The initial prior covariance is $\Xpriorinit = \begin{bmatrix}0.2 & 0.05\\ 0.05 & 0.1\end{bmatrix}$.
We use time-varying nominal covariances of the form $
\hat\Sigma_{w,t} = a_t\hat\Sigma_w$, $
\hat\Sigma_{v,t} = a_t\hat\Sigma_v
$, constructed by scaling the base matrices
\[
\hat\Sigma_w = \begin{bmatrix}0.3 & 0.1\\ 0.1 & 0.2\end{bmatrix},\quad
\hat\Sigma_v = \begin{bmatrix}0.15 & 0.05\\ 0.05 & 0.1\end{bmatrix},
\]
with piecewise-linear factors $a_t$:
\[
\begin{aligned}
a_t &= 
\begin{cases}
1 - 0.06\,t, & 0 \le t \le 5,\\
0.7 - 0.02(t-5), & 5 < t \le 15,\\
0.5, & t > 15.
\end{cases}
\end{aligned}
\]
These covariances are used at each step $t$ to generate the lower-bound KF (LOW-KF), upper-bound KF (HIGH-KF), and DRKF posterior covariance trajectories.  
To visualize the uncertainty, we use the $95\%$ confidence ellipse
$
\mathcal{E}(\Sigma) \coloneqq \{x \in \real{2}: x^\top \Sigma^{-1} x \leq \chi^2_{2,0.95}\}$ with $\chi^2_{2,0.95}=5.991$.
This region contains $95\%$ of a zero-mean Gaussian with covariance $\Sigma$.
By~\Cref{thm:drkf_sandwich},
$
\mathcal{E}(\Xpostlow)\subseteq \mathcal{E}(\Xpost)\subseteq \mathcal{E}(\Xposthigh)$.

As shown in~\Cref{fig:ellipses_2D}, the ellipses at representative time steps confirm the predicted nesting, providing a clear and interpretable robustness envelope.

\subsubsection{Convergence Property}\label{subsec:exp_converge}

\begin{figure}[t!]
    \centering
\includegraphics[width=0.7\linewidth]{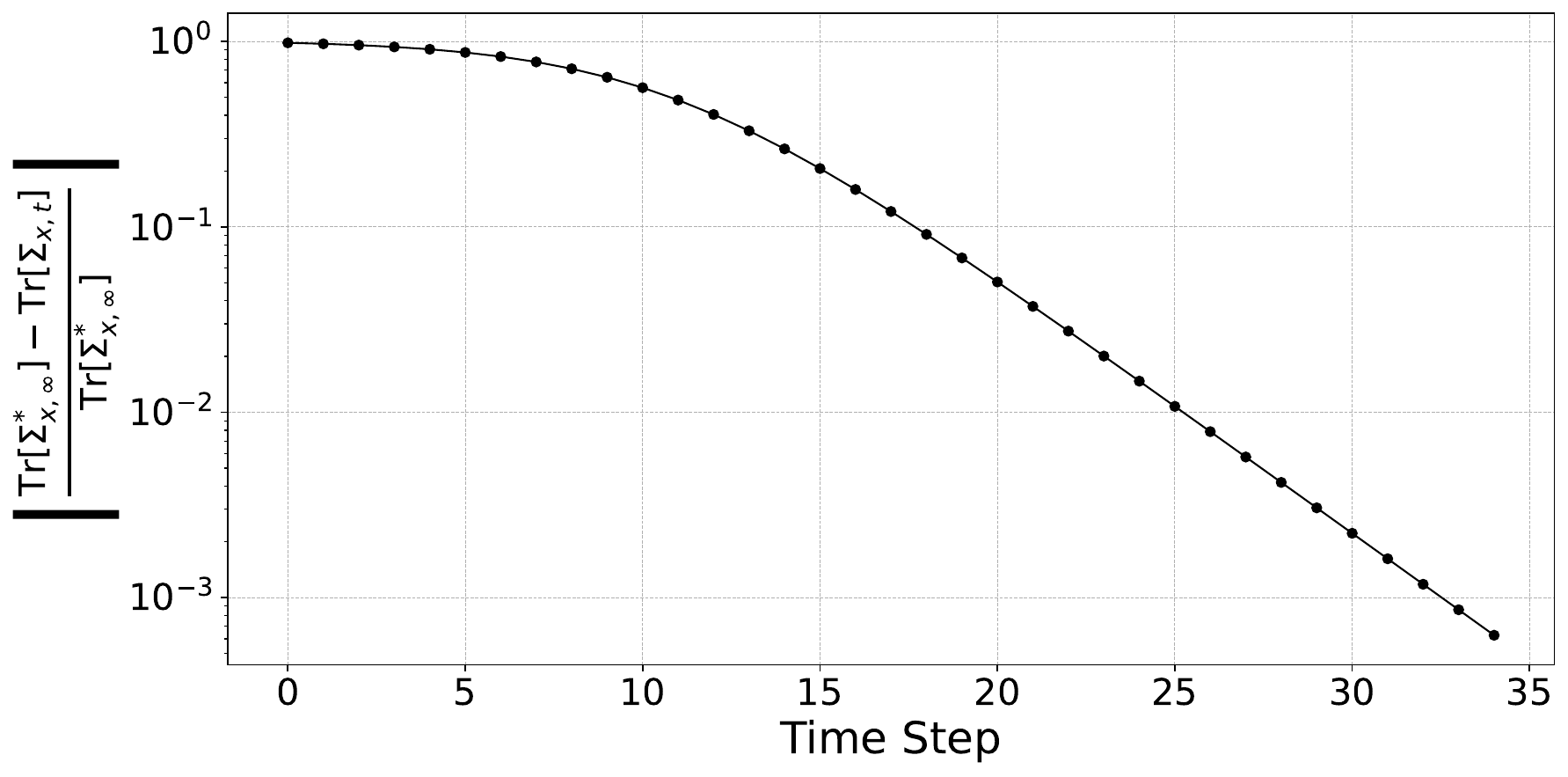}
    \caption{Relative difference between $\Tr[\Xpost]$ and $\Tr[\Xpostssopt]$.}
    \label{fig:convergence}
\end{figure}

Finally, we examine convergence of the time-varying DRKF to its steady-state counterpart using the
example from~\cite{zorzi2015convergence}:
$
A = \begin{bmatrix}
    0.1 & 1  \\ 1 & -1
\end{bmatrix}
, \quad C = \begin{bmatrix}
    1&-1
\end{bmatrix}$.\footnote{We set the nominal covariances to $\hat\Sigma_w = I_2$ and $\hat\Sigma_v = {1}$.} and the ambiguity set radii $\theta_{x_0}=\theta_w=\theta_v=0.1$.
\Cref{fig:convergence} shows that the relative error between $\Tr[\Xpost]$ and
$\Tr[\Xpostssopt]$ decays approximately linearly on a logarithmic scale, confirming exponential
convergence, consistent with the theoretical analysis.

\section{Conclusions}

We developed a DRKF based on a noise-centric Wasserstein ambiguity model. The proposed formulation preserves the structure of the classical KF, admits explicit spectral bounds, and ensures dynamic consistency under distributional uncertainty. 
We established existence, uniqueness, and convergence of the steady-state solution, proved asymptotic minimax optimality, and showed that the steady-state filter is obtained from a single stationary SDP with Kalman-level online complexity. 
Future directions include extending
the framework to nonlinear systems, developing adaptive mechanisms for online calibration of the
ambiguity set radius, incorporating model uncertainty in the system matrices, and exploring
temporally coupled ambiguity sets.

\appendix

\section{Eigenvalue Lower Bounds for DRSE Maximizers}\label{app:ev_lower}

\begin{lemma}\label{lem:ev_lower}
Consider the problems~\eqref{eqn:DRMMSE_init_opt} and~\eqref{eqn:DRMMSE_opt}
\emph{with the explicit eigenvalue lower-bound constraints removed}. Then:

For $t=0$, \eqref{eqn:DRMMSE_init_opt} admits a maximizer
$(\Xpriorinitopt,\Sigma_{v,0}^*)$ such that
\begin{align}\label{eqn:init_ev_ineq}
\Xpriorinitopt \succeq \lammin{\Xnom} I_{n_x}, \quad
\Sigma_{v,0}^* \succeq \lammin{\hat{\Sigma}_{v,0}} I_{n_y}.
\end{align}
For $t>0$, \eqref{eqn:DRMMSE_opt} admits a maximizer
$(\Sigma_{w,t-1}^*,\Sigma_{v,t}^*)$ such that
\begin{align}\label{eqn:ev_ineq}
\Sigma_{w,t-1}^* \succeq \lammin{\hat{\Sigma}_{w,t-1}} I_{n_x}, \quad
\Sigma_{v,t}^* \succeq \lammin{\hat{\Sigma}_{v,t}} I_{n_y}.
\end{align}
\end{lemma}
\begin{proof} \begin{sloppypar}
For $t=0$, \eqref{eqn:init_ev_ineq} is shown in \cite[Thm.~3.1]{nguyen2023bridging}.
We prove \eqref{eqn:ev_ineq} for $t>0$.
Fix $\Xpostprev$, so that $\Xprior=A_{t-1}\Xpostprev A_{t-1}^\top+\Sigma_{w,t-1}$.
Using the standard identity
$\Tr[\Xpost]
=
\inf_{K_t}\Tr \left[(I-K_tC_t)\Xprior(I-K_tC_t)^\top + K_t\Sigma_{v,t}K_t^\top\right]$, the objective can be written as
\begin{align*}
\Tr[\Xpost] =\inf_{K_t}\Big(
\langle (I-K_tC_t)^\top(I-K_tC_t),\, \Sigma_{w,t-1}\rangle  + f_t(K_t,\Sigma_{v,t})\Big),
\end{align*}
where  $\langle A,B\rangle \coloneqq \Tr[A^\top B]$ and $f_t$ collects the remaining terms (independent of $\Sigma_{w,t-1}$) and is convex and continuous in $K_t$.

Let $(\bar{\Sigma}_{w,t-1},\bar{\Sigma}_{v,t})$ be a maximizer of \eqref{eqn:DRMMSE_opt}  with the explicit eigenvalue bounds removed, which exists since the feasible set
(Bures--Wasserstein balls intersected with $\psd{\cdot}$) is compact and the objective is
continuous.
Fixing $\Sigma_{v,t}=\bar{\Sigma}_{v,t}$, the maximization over $\Sigma_{w,t-1}$ matches the
setting of \cite[Lemma~A.3]{nguyen2023bridging}, hence it admits a maximizer
$\Sigma_{w,t-1}^*\succeq \lammin{\hat{\Sigma}_{w,t-1}}I_{n_x}$.
Moreover, because $(\bar{\Sigma}_{w,t-1},\bar{\Sigma}_{v,t})$ is globally optimal,
the fixed-$\Sigma_{v,t}$ subproblem achieves the same optimal value, so
$(\Sigma_{w,t-1}^*,\bar{\Sigma}_{v,t})$ is also a global maximizer.

Now fix $\Sigma_{w,t-1}=\Sigma_{w,t-1}^*$, which in turn fixes
$\Xprior = A_{t-1}\Xpostprev A_{t-1}^\top + \Sigma_{w,t-1}^*$. 
Then, the dependence on
$\Sigma_{v,t}$ is of the form $\inf_{K_t}(\langle K_t^\top K_t,\Sigma_{v,t}\rangle + \tilde f_t(K_t))$,
so applying \cite[Lemma~A.3]{nguyen2023bridging} again yields a maximizer
$\Sigma_{v,t}^*\succeq \lammin{\hat{\Sigma}_{v,t}}I_{n_y}$, and
$(\Sigma_{w,t-1}^*,\Sigma_{v,t}^*)$ remains globally optimal. This proves \eqref{eqn:ev_ineq}.
\end{sloppypar}
\end{proof}

\section{Proofs for Spectral Boundedness}

\subsection{Proof of~\Cref{prop:bures_bound}} \label{app:bures_bound}
By~\Cref{lem:procr}, there exists $U \in O(n)$ such that
$
\|X^{1/2} - \hat{X}^{1/2}U \|_F \leq \theta.
$
Since $\| \cdot \|_2 \leq \|\cdot\|_F$, we also have
\begin{align}\label{eqn:norm2_thetax}
\|X^{1/2} - \hat{X}^{1/2}U \|_2 \leq \theta.
\end{align}
For any matrices $A,B \in \real{n\times n}$, the singular-value perturbation bounds~\cite[Cor.~7.3.5]{horn2012matrix} state that
\begin{equation}
\begin{split} \label{eqn:sig_per}
    &|\sigma_{\max}(A) - \sigma_{\max}(B) | \leq \| A - B\|_2 \\
    &|\sigma_{\min}(A) - \sigma_{\min}(B)| \leq \|A-B\|_2.
\end{split}
\end{equation}
Apply~\eqref{eqn:sig_per} with $A = X^{1/2}$ and $B = \hat{X}^{1/2}U$, and combine with~\eqref{eqn:norm2_thetax} to obtain $|\sigma_{\max}(X^{1/2}) - \sigma_{\max}(\hat{X}^{1/2}U) | \leq \theta$ and $|\sigma_{\min}(\hat{X}^{1/2}U) - \sigma_{\min}(X^{1/2}) |\leq  \theta$.
Because $U$ is orthogonal, $X \succeq 0$ and $\hat{X} \succeq 0$, we have 
$\sigma_{\max}(\hat{X}^{1/2}U)= \sigma_{\max}(\hat{X}^{1/2}) = \sqrt{\lammax{\hat{X}}}$ and $\sigma_{\min}(\hat{X}^{1/2}U) = \sigma_{\min}(\hat{X}^{1/2}) = \sqrt{\lammin{\hat{X}}}$.
Substituting these into the inequalities above yields
   \[
   \begin{split}
   \sqrt{\lammax{X}}&=\sigma_{\max}(X^{1/2})\leq\sqrt{\lammax{\hat{X}}}+\theta,  \\
   \sqrt{\lammin{X}}&=\sigma_{\min}(X^{1/2})\geq\sqrt{\lammin{\hat{X}}}-\theta.
   \end{split}
   \]
Squaring both sides and clipping the lower bound at zero yields the claim.

\subsection{Proof of~\Cref{cor:bounds}}\label{app:bounds}

By~\Cref{lem:GelbrichMMSE}, the matrices
$\Sigma_{v,t}^*, \Sigma_{w,t}^*$ and $\Xpriorinitopt$ belong to the Bures--Wasserstein balls centered at $\hat{\Sigma}_{v,t}, \hat{\Sigma}_{w,t}$ and $\Xnom$ with radii $\theta_v,\theta_w$ and $\theta_{x_0}$, respectively. Hence,~\Cref{prop:bures_bound} gives the corresponding upper bounds.

The lower bounds follow directly from the constraints  $\Sigma_{v,t} \succeq \lammin{\hat{\Sigma}_{v,t}} I_{n_y}, \Sigma_{w,t-1} \succeq \lammin{\hat{\Sigma}_{w,t-1}} I_{n_x}$, and $ \Xpriorinit \succeq \lammin{\Xnom} I_{n_x}$, which are imposed in~\eqref{eqn:DRMMSE_init_opt} and~\eqref{eqn:DRMMSE_opt}.
Thus, any maximizers $\Sigma_{v,t}^*$, $\Sigma_{w,t-1}^*$, and $\Xpriorinitopt$ necessarily satisfy these inequalities, giving the claimed lower bounds.

\subsection{Proof of~\Cref{thm:drkf_sandwich}}\label{app:drkf_sandwich}

Define the posterior and prediction maps as
\[
\begin{split}
\Psi(\Xprior,\Sigma_{v,t}) &\coloneqq \Xprior - \Xprior C_t^{\top}(C_t \Xprior C_t^{\top}+ \Sigma_{v,t})^{-1} C_t \Xprior\\
\Phi (\Xpost , \Sigma_{w,t}) &\coloneqq A_t \Xpost A_t^{\top} + \Sigma_{w,t}.
\end{split}
\]
The DRKF recursion is obtained by evaluating these maps at the least-favorable noise covariances:
$
\Xpost=\Psi(\Xprior,\Sigma_{v,t}^{*})$ and $
\Xpriornext=\Phi(\Xpost,\Sigma_{w,t}^{*})$. 
For $\Sigma_{v,t}\succ 0$, the posterior covariance admits the variational
representation
\[
\Psi(\Xprior,\Sigma_{v,t})
=
\inf_{K_t}\Big[(I-K_tC_t)\Xprior(I-K_tC_t)^{\top}
+ K_t\Sigma_{v,t}K_t^{\top}\Big].
\]
Hence, $\Psi$ is monotone increasing in each argument: if
${\Xprior}^{(1)}\succeq {\Xprior}^{(2)}$ and
$\Sigma_{v,t}^{(1)}\succeq \Sigma_{v,t}^{(2)}$, then for any $K_t$ the corresponding
quadratic form is larger, and taking the infimum preserves the Loewner order.
 The map $\Phi$ is also monotone since congruence with $A_t$ and addition of a PSD matrix preserve the Loewner order.

\Cref{cor:bounds}, together with the monotonicity of the posterior and prediction maps $\Psi$ and $\Phi$, implies by induction that, for all $t\ge 0$, the KF driven by $(\underline{\lambda}_{w,t} I_{n_x},\underline{\lambda}_{v,t} I_{n_y})$ produces covariances no larger than those of the DRKF, and the KF driven by $(\overline{\lambda}_{w,t} I_{n_x},\overline{\lambda}_{v,t} I_{n_y})$ produces covariances no smaller.

\section{Proofs for Convergence}
\subsection{Proof of~\Cref{lem:WO_pd}} \label{app:WO_pd}

By~\Cref{cor:bounds} and~\Cref{assump:time_inv_nom}, there exist constants $\underline{\lambda}_w, \underline{\lambda}_v > 0$ such that $\Sigma_{w,t}^*\succeq \underline{\lambda}_w I_{n_x} \succ 0$ and $\Sigma_{v,t}^*\succeq \underline{\lambda}_v I_{n_y} \succ 0$. This implies $\Sigma_{w,t}^{N,*} \succ 0$ and $\Sigma_{v,t}^{N,*} \succ 0$, and consequently, 
\[
\Sigma_{u,t}^N=\mathcal{H}_N\Sigma_{w,t}^{N,*}\mathcal{H}_N^\top+\Sigma_{v,t}^{N,*}\succ0.
\] 

Using the matrix inversion lemma, the conditional covariance $\mathcal{Q}_{N,t}$ admits the equivalent representation
\[\mathcal{Q}_{N,t} = \big((\Sigma_{w,t}^{N,*})^{-1} + \mathcal{H}_N^\top (\Sigma_{v,t}^{N,*})^{-1} \mathcal{H}_N\big)^{-1}.\] Since $(\Sigma_{w,t}^{N,*})^{-1} \succ 0$ and $\mathcal{H}_N^\top (\Sigma_{v,t}^{N,*})^{-1} \mathcal{H}_N \succeq 0$, it follows that $\mathcal{Q}_{N,t} \succ 0$.
Moreover, $\mathcal{R}_N$ has full row rank for all $N \geq 1$. Hence, for any nonzero $x\in\real{n_x}$,
\[x^\top W_{N,t} x = (\mathcal{R}_N^\top x)^\top \mathcal{Q}_{N,t} (\mathcal{R}_N^\top x) > 0,\]
which proves that $W_{N,t} \succ 0$.

Finally, since $\Sigma_{u,t}^N \succ 0$, we have $(\Sigma_{u,t}^{N})^{-1}\succ 0$. By~\Cref{assump:obsv}, the observability of $(A,C)$ implies that $\mathcal{O}_{N}$ has full column rank for any $N \geq n_x$. Therefore, for any nonzero $x \in \real{n_x}$,
\[x^\top \Omega_{N,t} x = (\mathcal{O}_N x)^\top (\Sigma_{u,t}^N)^{-1}(\mathcal{O}_N x) > 0,\]
which establishes $\Omega_{N,t} \succ 0$ for all $N\geq n_x$.

\subsection{Proof of~\Cref{prop:xi_bound}}\label{app:xi_bound}

Fix $\theta_w$ and $\theta_v$ and let
$\underline{\lambda}_w \coloneq \lammin{\hat{\Sigma}_w}$,
$\underline{\lambda}_v \coloneq \lammin{\hat{\Sigma}_v}$,
$\overline{\lambda}_w \coloneq \overline{\lambda}(\hat{\Sigma}_w,\theta_w)$, and
$\overline{\lambda}_v \coloneq \overline{\lambda}(\hat{\Sigma}_v,\theta_v)$ as in~\Cref{cor:bounds}. Then, the least-favorable covariances satisfy
$\underline{\lambda}_w I \preceq \Sigma_{w,t}^{N,*} \preceq \overline{\lambda}_w I$ and $\underline{\lambda}_v I \preceq \Sigma_{v,t}^{N,*} \preceq \overline{\lambda}_v I$.
Hence,
\[
\Sigma_{u,t}^N
=
\mathcal{H}_N \Sigma_{w,t}^{N,*} \mathcal{H}_N^\top + \Sigma_{v,t}^{N,*}
\preceq
(\overline{\lambda}_w \|\mathcal{H}_N\|_2^2 + \overline{\lambda}_v) I,
\]
and $\Sigma_{u,t}^N \succeq \underline{\lambda}_v I$.
Therefore,
\[
\lammin{\Omega_{N,t}}
\geq
\frac{\sigma_{\min}(\mathcal{O}_N)^2}{\lammax{\Sigma_{u,t}^N}}
\geq
\frac{\sigma_{\min}(\mathcal{O}_N)^2}{\overline{\lambda}_v + \overline{\lambda}_w \|\mathcal{H}_N\|_2^2},
\]
which implies
\[
\|\Omega_{N,t}^{-1}\|_2
\le
\frac{\overline{\lambda}_v + \overline{\lambda}_w \|\mathcal{H}_N\|_2^2}{\sigma_{\min}(\mathcal{O}_N)^2}.
\]
Next, using the definition of $\mathcal{G}_{N,t}$, we obtain
\[
\|\mathcal{G}_{N,t}\|_2
\le
\|\Sigma_{w,t}^{N,*}\|_2 \|\mathcal{H}_N\|_2 \, \|(\Sigma_{u,t}^N)^{-1}\|_2
\leq
\frac{\overline{\lambda}_w}{\underline{\lambda}_v} \|\mathcal{H}_N\|_2.
\]
Consequently,
\[
\begin{split}
\|\alpha_{N,t}\|_2
&\leq
\|A^N\|_2
+
\|\mathcal{R}_N\|_2 \|\mathcal{G}_{N,t}\|_2 \|\mathcal{O}_N\|_2 \\
&\leq
\|A^N\|_2
+
\frac{\overline{\lambda}_w}{\underline{\lambda}_v}
\|\mathcal{R}_N\|_2 \|\mathcal{H}_N\|_2
\|\mathcal{O}_N\|_2.
\end{split}
\]
Finally, we have that
\[
\begin{split}
\lammin{\mathcal{Q}_{N,t}}
&\ge
\frac{1}{
\lammax{(\Sigma_{w,t}^{N,*})^{-1}}
+
\lammax{\mathcal{H}_N^\top (\Sigma_{v,t}^{N,*})^{-1} \mathcal{H}_N}
} \\
&\geq
\frac{1}{
1/\underline{\lambda}_w
+
\|\mathcal{H}_N\|_2^2/\underline{\lambda}_v
}.
\end{split}
\]
Since $\mathcal{R}_N \mathcal{R}_N^\top \succeq I$, we have
$\lammin{W_{N,t}} \geq \lammin{\mathcal{Q}_{N,t}}$,
and therefore 
\[
\|W_{N,t}^{-1}\|_2
\leq
1/\underline{\lambda}_w
+
\|\mathcal{H}_N\|_2^2/\underline{\lambda}_v.
\]
Combining the above bounds yields
\[
\|M_{N,t}\|_2
\leq
\|\Omega_{N,t}^{-1}\|_2 \|\alpha_{N,t}\|_2^2 \|W_{N,t}^{-1}\|_2
\leq
\kappa_N(\theta_v,\theta_w).
\]
Substituting into~\eqref{eqn:contr_factor} and using the monotonicity of
$s\mapsto (\frac{\sqrt{s}}{1+\sqrt{1+s}})^2$ on $s\ge 0$ gives
$\xi_{N,t}\le\bar\xi_N(\theta_v,\theta_w)$; inequality~\eqref{eqn:rho_to_kappa}
then ensures $\xi_{N,t}\le\rho$ for all $t\ge 0$.

\subsection{Proof of~\Cref{thm:contraction}} \label{app:contraction}

Fix $t \geq 0$ and $N \geq n_x$. By~\Cref{lem:WO_pd}, we have $W_{N,t} \succ 0$ and $\Omega_{N,t} \succ 0$. Hence, the downsampled DR Riccati mapping $r_{\ambset,t}^d$ in~\eqref{eqn:downsampled_Riccati} is well-defined on $\pd{n_x}$ and has the robust Riccati form considered in~\cite[Thm.~5.3]{lee2008invariant}. Therefore, $r_{\ambset,t}^d$ is a strict contraction with respect to Thompson's part metric on $\pd{n_x}$, with contraction factor $\xi_{N,t}\in(0,1)$ satisfying the bound~\eqref{eqn:contr_factor}.

By~\Cref{prop:xi_bound}, we have $\xi_{N,t}\le \bar\xi_N(\theta_v,\theta_w)<1$
for all $t\ge0$, and thus $r_{\ambset,t}^d$ is uniformly contractive.
Under~\Cref{assump:time_inv_nom}, the one-step DR Riccati update is stationary,
so the downsampled map coincides with the $N$-fold iterate of a single
time-invariant map.
 The contraction argument for downsampled iterations
in~\cite{zorzi2017convergence} then implies that the one-step DR Riccati
recursion~\eqref{eqn:DR_Riccati_def} converges to a unique fixed point
$\Xpriorss\succ0$.

Finally, although the stage-wise SDP may admit multiple optimizers in
$\Sigma_{v,t}^*$, the corresponding DR Kalman gain  is unique.
By \Cref{lem:GelbrichMMSE}, for a fixed $\Xprior$ the stage-wise DRSE problem
amounts to minimizing over $K$ the worst-case posterior MSE, with the adversary
maximizing over $\Sigma_{v,t}$ in the SDP's feasible set, denoted by $\mathcal A_{v, t}$. Using the standard
KF covariance identity, the resulting worst-case one-step MSE is
\[
\sup_{\Sigma_{v,t}\in\mathcal A_{v, t}}
\Tr\big[(I-KC)\Xprior(I-KC)^\top + K\Sigma_{v,t}K^\top\big].
\]
For each feasible $\Sigma_{v,t}$, the associated quadratic map
\[
K \mapsto 
\Tr\big[(I-KC)\Xprior(I-KC)^\top + K\Sigma_{v,t}K^\top\big]
\]
is $\underline{\lambda}_v$-strongly convex in $K$ (with respect to
$\|\cdot\|_F$).
Taking the supremum  over the compact feasible set $\mathcal A_{v,t}$ preserves
strong convexity, and hence the minimizer $K_t$ is unique.
Moreover, this worst-case objective is continuous in
$(\Xprior,K)$,
 so the unique minimizer $K_t$ depends continuously on $\Xprior$.
Therefore, $\Xprior\to \Xpriorss$ implies that the sequence $\{K_t\}$ converges.

\subsection{Proof of \Cref{cor:ss_sdp}}\label{app:ss_sdp}

For each $t\ge 1$, let
$z_t := (\Xprior,\Xpost,\Sigma_{w,t-1}^*,\Sigma_{v,t}^*,Y_t,Z_t)$
denote a primal optimal solution of the stage-wise SDP~\eqref{eqn:DRKF_SDP}.
By \Cref{thm:contraction}, $(\Xprior,\Xpost)\to(\Xpriorss,\Xpostss)$.

Using the LMIs
$\begin{bmatrix}
\hat{\Sigma}_{w} & Y_t\\
Y_t^\top & \Sigma_{w,t-1}^*
\end{bmatrix}\succeq 0$ and
$\begin{bmatrix}\hat{\Sigma}_{v} & Z_t\\
Z_t^\top & \Sigma_{v,t}^*
\end{bmatrix}\succeq 0$,
together with $\hat{\Sigma}_{w}\succ 0$, $\hat{\Sigma}_{v}\succ 0$, 
Schur complements yield
$\Sigma_{w,t-1}^* \succeq Y_t^\top \hat\Sigma_w^{-1} Y_t$ and
$\Sigma_{v,t}^* \succeq Z_t^\top \hat\Sigma_v^{-1} Z_t$.
Hence, it follows from \Cref{cor:bounds} that
\[\|Y_t\|_2^2 \le \|\hat\Sigma_w\|_2\ \|\Sigma_{w,t-1}^*\|_2 \le \|\hat\Sigma_w\|_2\ \bar\lambda_w\]
and
\[\|Z_t\|_2^2 \le \|\hat\Sigma_v\|_2\ \|\Sigma_{v,t}^*\|_2 \le \|\hat\Sigma_v\|_2\ \bar\lambda_v,\]
so $\{Y_t\}$ and $\{Z_t\}$ are uniformly bounded.
Therefore, $\{z_t\}$ is bounded, and by the Bolzano--Weierstrass theorem,
there exists a subsequence $t_k$ such that $z_{t_k}\to z_\infty$.
Writing \[
z_\infty :=
(\Xpriorss,\Xpostss,\Sigma_{w,\infty}^*,\Sigma_{v,\infty}^*,Y_\infty,Z_\infty),\]
we have $(\Sigma_{w,t_k-1}^*,\Sigma_{v,t_k}^*)\to (\Sigma_{w,\infty}^*,\Sigma_{v,\infty}^*)$
along this subsequence.

Passing to the limit along this subsequence in the constraints of
\eqref{eqn:DRKF_SDP}, and using $\Sigma_{x, t_k-1}\to\Xpostss$  and the closedness of the PSD cone, shows that
$z_\infty$
is feasible for the stationary SDP~\eqref{eqn:DRKF_SDP_ss}.

We now show that $z_\infty$ is in fact \emph{optimal} for \eqref{eqn:DRKF_SDP_ss}.
Assume $\theta_w>0$ and $\theta_v>0$.\footnote{The degenerate case
$\theta_w=\theta_v=0$ reduces to the nominal KF, for which the statement is
immediate.}
Then \eqref{eqn:DRKF_SDP_ss} satisfies Slater's condition.
For instance, choose
$\Sigma_{w,\infty}=\hat\Sigma_w+\varepsilon I$, $\Sigma_{v,\infty}=\hat\Sigma_v+\varepsilon I$,
$Y=\hat\Sigma_w$, $Z=\hat\Sigma_v$, $\Xpostss=0$, and $\Xpriorss=\Sigma_{w,\infty}$
for sufficiently small $\varepsilon>0$; this makes all LMIs $\succ 0$ and the trace
inequalities strict.
Therefore, strong duality holds and the KKT conditions are necessary and sufficient
for optimality.

For each $t\ge1$, let $\lambda_t$ be a dual optimal solution associated with the
primal optimizer $z_t$ of \eqref{eqn:DRKF_SDP} (existence follows from Slater's theorem).
Moreover, since the primal feasible sets are uniformly bounded and Slater's condition holds,
the set of primal--dual optimal pairs is bounded.
Thus, after passing to a further subsequence if necessary, we have
$\lambda_{t_k}\to \lambda_\infty$.
Taking limits in the KKT conditions of \eqref{eqn:DRKF_SDP} along $t_k$ (all
primal/dual feasibility constraints are closed, and complementarity is preserved
under limits) yields that $(z_\infty,\lambda_\infty)$ satisfies the KKT
conditions of the stationary SDP \eqref{eqn:DRKF_SDP_ss}.
Consequently, $z_\infty$ is optimal for~\eqref{eqn:DRKF_SDP_ss}, and
$(\Xpriorss,\Xpostss)=(\Xpriorssopt,\Xpostssopt)$.

Finally, by \Cref{thm:contraction}, the gain sequence $\{K_t\}$ converges.
Along the subsequence $t_k$, 
$(\Sigma_{x,t_k}^-,\Sigma_{v,t_k}^*)\to(\Xpriorssopt,\Sigma_{v,\infty}^*)$, and
continuity of
$K(\Xprior,\Sigma_v)
=\Xprior C^\top(C\Xprior C^\top+\Sigma_v)^{-1}$
on $\pd{n_x}\times\pd{n_y}$ yields
$\lim_{k\to\infty}K_{t_k}=K_\infty^*$.
Since $\{K_t\}$ has a unique limit, we conclude that
$K_\infty^*=\lim_{t\to\infty}K_t$.

\section{Proofs for Stability and Optimality}

\subsection{Proof of~\Cref{thm:drkf_stability}}\label{app:drkf_stability}

Let $\tilde{e}_t := x_t - \xprior_t$ denote the \emph{prediction} error.
From the DRKF update,
\[
e_t = x_t - \xpost_t = (I-K_\infty^* C)\,\tilde e_t - K_\infty^* (v_t - \hat{v}).\]
The prediction error evolves as
$\tilde{e}_t = A e_{t-1} + w_{t-1}-\hat{w}$.
Eliminating $\tilde{e}_t$ yields the posterior error recursion
\begin{align}\label{eqn:error_recursion}
\begin{split}e_t &= (I-K_\infty^* C)A\,e_{t-1} + (I-K_\infty^* C)\,(w_{t-1} - \hat{w}) - K_\infty^* (v_t - \hat{v})
\end{split}\\
&= F_\infty e_{t-1} + (I-K_\infty^* C)(w_{t-1}-\hat{w}) - K_\infty^* (v_t-\hat{v}).\nonumber
\end{align}

At steady-state, the DRKF covariances satisfy
\[
\Xpriorssopt = A \Xpostssopt A^\top + \Sigma_{w,\infty}^*\]
with $K_\infty^* = \Xpriorssopt C^\top\!\bigl(C\,\Xpriorssopt C^\top+\Sigma_{v,\infty}^*\bigr)^{-1}$.
Using the Joseph form,
\[
\Xpostssopt \;=\; (I-K_\infty^* C)\,\Xpriorssopt\,(I-K_\infty^* C)^\top + K_\infty^* \Sigma_{v,\infty}^* (K_\infty^*)^\top.
\]
Substituting $\Xpriorssopt = A \Xpostssopt A^\top + \Sigma_{w,\infty}^*$ gives
\begin{align*}
\Xpostssopt
= F_\infty\,\Xpostssopt\,F_\infty^\top
   + (I-K_\infty^* C)\,\Sigma_{w,\infty}^*\,(I-K_\infty^* C)^\top + K_\infty^* \Sigma_{v,\infty}^* (K_\infty^*)^\top.
\end{align*}
Hence,
\[
\Xpostssopt  -  F_\infty \Xpostssopt F_\infty^\top = Q_\infty,
\]
where $Q_\infty
:= (I-K_\infty^* C)\Sigma_{w,\infty}^*(I-K_\infty^* C)^\top
   + K_\infty^*\Sigma_{v,\infty}^*(K_\infty^*)^\top$.
Since $\Sigma_{w,\infty}^*,\Sigma_{v,\infty}^* \succ 0$, for any $x\neq0$,
\[
x^\top Q_\infty x
= \|\Sigma_{w,\infty}^{*,1/2}(I-K_\infty^* C)^\top x\|^2
  + \|\Sigma_{v,\infty}^{*,1/2}(K_\infty^*)^\top x\|^2 > 0,
\]
which implies $Q_\infty\succ0$. Thus,
$\Xpostssopt \succ F_\infty \Xpostssopt F_\infty^\top$.
By the standard discrete-time Lyapunov characterization, this strict inequality
implies that $F_\infty$ is Schur stable.

Using the temporal independence of $\{w_t\}$ and $\{v_t\}$ and their independence
from $e_{t-1}$, the error covariance $P_t:=\mathrm{cov}(e_t)$ satisfies
\[
P_t = F_\infty P_{t-1}F_\infty^\top + Q_\infty,
\]
whose unique fixed point is $\Xpostssopt$.
Since $F_\infty$ is Schur, $P_t\to\Xpostssopt$ and
$\sup_{t\ge0}\mathbb E[\|e_t\|^2]<\infty$, completing the proof.

\subsection{Proof of~\Cref{cor:drkf_error_bias}}\label{app:drkf_error_bias}

Taking expectations in~\eqref{eqn:error_recursion} yields
$m_t \;=\; F_\infty m_{t-1} + (I-K_\infty^* C)(\mu_w-\hat{w}) - K_\infty^* (\mu_v-\hat{v})$.
Since $F_\infty$ is Schur by~\Cref{thm:drkf_stability}, 
we have $\rho(F_\infty)<1$, and hence $I-F_\infty$ is invertible with
$(I-F_\infty)^{-1}=\sum_{k=0}^\infty F_\infty^k$.
Therefore, the affine recursion converges, yielding
\[
\lim_{t\to\infty} m_t
= (I - F_\infty)^{-1} \bigl((I-K_\infty^* C)(\mu_w-\hat{w}) - K_\infty^* (\mu_v-\hat{v})\bigr).
\]
If $\hat{w}=\mu_w$ and $\hat{v}=\mu_v$, the constant input term vanishes, and thus $m_t\to 0$.

\subsection{Proof of~\Cref{thm:optim}}\label{app:optim}

For each $t\ge 0$, define the stage-wise minimax value  
\[
V_t \coloneqq \inf_{\psi_t\in\mathcal{F}_t} \sup_{\Pdist_{e,t}\in\ambset_{e,t}}
J_t(\psi_t,\Pdist_{e,t}),
\]
where $J_t(\psi_t,\Pdist_{e,t})$ is understood almost surely with respect to
$\sigma(\mathcal{Y}_{t-1})$.
Under~\Cref{assump:Gauss}, \Cref{lem:GelbrichMMSE} and \Cref{thm:DRKF} imply that the
minimax problem admits a Gaussian least-favorable distribution and that the
corresponding optimal estimator is the DRKF $\psi_t^*$. Moreover, the minimax
value is given by
$V_t = \Tr[\Xpost]$,
where $\Xpost = \Psi(\Xprior,\Sigma_{v,t}^*)$ is the DRKF posterior  covariance at time $t$.
Consequently, for any admissible estimator $\psi_t\in\mathcal{F}_t$,
\[
\sup_{\Pdist_{e,t}\in\ambset_{e,t}} J_t(\psi_t,\Pdist_{e,t})
\ge 
V_t
=
\Tr[\Xpost].
\]
Taking $\liminf_{t\to\infty}$ and using \Cref{thm:contraction}, which guarantees
$\Xpost\to\Xpostssopt$, yields~\eqref{eqn:lower_bound_liminf}.
The corresponding long-run average lower bound~\eqref{eqn:lower_bound_avg} follows by Ces\`aro summability.

We now prove the asymptotic optimality of the steady-state DRKF $\psi_\infty$ with gain $K_\infty^*$.
Let $\xpost_t^\infty:=\psi_\infty(\mathcal{Y}_t)$ and
$\xpost_t^*:=\psi_t^*(\mathcal{Y}_t)$ denote the posterior estimates produced by the
steady-state and time-varying DRKFs, respectively.
Define the difference $d_t:=\xpost_t^\infty-\xpost_t^*$.

A direct comparison of the measurement updates yields
\begin{equation}\label{eq:d_rec_appopt}
d_t = F_\infty d_{t-1} + \Delta_t \eta_t^*, \quad t\ge 1,
\end{equation}
where $F_\infty:=(I-K_\infty^*C)A$, $\Delta_t:=K_\infty^*-K_t$, and
$\eta_t^*:=y_t-C {\xprior_t}^*-\hat v$ is the innovation of the time-varying DRKF.
By \Cref{thm:drkf_stability}, $F_\infty$ is Schur, and by
\Cref{thm:contraction} we have $\|\Delta_t\|_2\to 0$.

Next, we bound the innovation second moment uniformly.
Since $\eta_t^* = C(x_t-{\xprior_t}^*) + (v_t-\hat v)$,
\[
\begin{split}
\mathbb{E} \left[\|\eta_t^*\|^2 \mid \mathcal{Y}_{t-1}\right]
\le
2\|C\|_2^2\,\mathbb{E}\left[\|x_t-{\xprior_t}^*\|^2\mid\mathcal{Y}_{t-1}\right] +
2\,\mathbb{E}\left[\|v_t-\hat v\|^2\mid\mathcal{Y}_{t-1}\right].
\end{split}
\]
Under \Cref{assump:time_inv_nom}, the ambiguity sets have finite Wasserstein
radii, so all $\Pdist_{v,t}\in\ambset_{v,t}$ have uniformly bounded second
moments about $\hat v$.
Moreover, \Cref{thm:contraction} implies $\Xpost\to\Xpostssopt$, and hence
$\sup_{t\ge0}\Tr[\Xpost]<\infty$.
Since $\Xprior = A\Sigma_{t-1}A^\top + \Sigma_{w,t-1}^*$ and
$\Sigma_{w,t-1}^*\preceq \overline\lambda_w I$ by \Cref{cor:bounds}, it follows
that $\sup_{t\ge0}\Tr[\Xprior]<\infty$, and therefore
\[
\sup_{t\ge0}\mathbb{E}[\|x_t-{\xprior_t}^*\|^2\mid\mathcal{Y}_{t-1}]<\infty.
\]
Consequently, there exists $M_\eta<\infty$ such that
\[
\sup_{t\ge 0}\ \sup_{\Pdist_{e,t}\in\ambset_{e,t}}
\mathbb{E} \left[\|\eta_t^*\|^2 \mid \mathcal{Y}_{t-1}\right]
\le M_\eta.
\]
Therefore,
\[
\sup_{\Pdist_{e,t}\in\ambset_{e,t}}
\mathbb{E}\left[\|\Delta_t\eta_t^*\|^2 \mid \mathcal{Y}_{t-1}\right]
\le \|\Delta_t\|_2^2\,M_\eta
\xrightarrow [t\to\infty]{} 0.
\]
Since $F_\infty$ is Schur, iterating~\eqref{eq:d_rec_appopt} and using the above
uniform bound yields that $d_t\to 0$ in mean square; in particular,
\begin{equation*}\label{eq:d_to_0_appopt}
D_t := \sup_{\Pdist_{e,t}\in\ambset_{e,t}}
\mathbb{E} \left[\|d_t\|^2 \mid \mathcal{Y}_{t-1}\right]
\xrightarrow[t\to\infty]{} 0.
\end{equation*}

Let $e_t^*:=x_t-\xpost_t^*$ denote the time-varying DRKF posterior error.
Since $x_t-\xpost_t^\infty = e_t^*-d_t$, we have
\[
\|x_t-\xpost_t^\infty\|^2
\le \|e_t^*\|^2 + \|d_t\|^2 + 2\|e_t^*\| \|d_t\|.
\]
Taking conditional expectations and then the supremum over $\ambset_{e,t}$ yields
\[
\begin{split}
\sup_{\Pdist_{e,t}\in\ambset_{e,t}}J_t(\psi_\infty,\Pdist_{e,t})
\le\sup_{\Pdist_{e,t}\in\ambset_{e,t}}J_t(\psi_t^*,\Pdist_{e,t})
+ D_t
+ 2\sqrt{D_t}\sqrt{\sup_{\Pdist_{e,t}\in\ambset_{e,t}}J_t(\psi_t^*,\Pdist_{e,t})}.
\end{split}
\]
By stage-wise minimax optimality of the time-varying DRKF,
$\sup_{\Pdist_{e,t}\in\ambset_{e,t}}J_t(\psi_t^*,\Pdist_{e,t})=V_t=\Tr[\Xpost]$.
Thus,
\begin{equation}\label{eq:ub_gap_appopt}
\sup_{\Pdist_{e,t}\in\ambset_{e,t}}J_t(\psi_\infty,\Pdist_{e,t})
\le
V_t + D_t + 2\sqrt{V_t}\sqrt{D_t}.
\end{equation}
Taking $\limsup_{t\to\infty}$ and using $D_t\to0$ and
$V_t\to \Tr[\Xpostssopt]$ gives
\[
\limsup_{t\to\infty}
\sup_{\Pdist_{e,t}\in\ambset_{e,t}}J_t(\psi_\infty,\Pdist_{e,t})
\le \Tr[\Xpostssopt].
\]
Combined with the lower bound~\eqref{eqn:lower_bound_liminf} applied to
$\psi_\infty$, this proves \eqref{eq:psi_infty_stage}.
The Ces\`aro-average optimality statement \eqref{eq:psi_infty_cesaro}  follows similarly since  
$\sup_t V_t<\infty$ and $D_t\to0$ imply $\frac1T\sum_{t=0}^{T-1}D_t\to0$,
and hence all error terms in \eqref{eq:ub_gap_appopt} have vanishing Ces\`aro mean.

\bibliographystyle{IEEEtran}

\bibliography{ref}

 \vspace{-0.15in}
\end{document}